%% file: main.tex
\documentclass[12pt,a4paper]{article}

\usepackage{ifthen} 
\usepackage{comment}
\newboolean{pdflatex}
\setboolean{pdflatex}{true} 

\newboolean{articletitles}
\setboolean{articletitles}{true} 

\newboolean{uprightparticles}
\setboolean{uprightparticles}{false} 

\newboolean{inbibliography}
\setboolean{inbibliography}{false} 

\def\paperauthors{LHCb Collaboration} 
\def\paperasciititle{Observation of a JPsiLz resonance consistent with a strange pentaquark candidate in BmToJpsiLzAnti-p decays} 
\def\papertitle{Observation of a \jpsi\Lz resonance consistent with a strange pentaquark candidate in \decay{\Bm}{\jpsi\Lz\antiproton} decays} 
\def\paperkeywords{{High Energy Physics}, {LHCb}} 
\def\papercopyright{\the\year\ CERN for the benefit of the LHCb collaboration} 
\def\paperlicence{CC BY 4.0 licence}
\def\paperlicenceurl{https://creativecommons.org/licenses/by/4.0/}

\input{preamble}
\usepackage{longtable} 
\begin{document}

\renewcommand{\thefootnote}{\fnsymbol{footnote}}
\setcounter{footnote}{1}
\newcommand{\HellH}[1]{\ensuremath{H^{#1}}}
\newcommand{\wigner}{\zeta}
\def\LbarTopbarpip {\decay{\Lbar}{\antiproton\pip}}
\def\Pcs {} 
\def\@processeq$#1\pm#2$\relax{\def\lhs{#1}\def\rhs{#2}}
\def\PL      {
{\ensuremath{P_{\psi s}^{\Lz}^0}\xspace}}
\def\PNm      {
{\ensuremath{\overline{P}_{\psi }^{N}^-}\xspace}}
\def\PNp      {
{\ensuremath{P_{\psi }^{N}^+}\xspace}}
\input{title-LHCb-PAPER}


\renewcommand{\thefootnote}{\arabic{footnote}}
\setcounter{footnote}{0}

\cleardoublepage


\pagestyle{plain} 
\setcounter{page}{1}
\pagenumbering{arabic}


\input{body}

\input{acknowledgements}


\clearpage
\addcontentsline{toc}{section}{References}
\setboolean{inbibliography}{true}
\bibliographystyle{LHCb}
\bibliography{main,standard,LHCb-PAPER,LHCb-CONF,LHCb-DP,LHCb-TDR}
 
\clearpage
 \newpage
\input{appendix}
\clearpage
 
\newpage
\input{Authorship_LHCb-PAPER-2022-031}


\end{document}

%% file: preamble.tex
\usepackage[top=1in, bottom=1.25in, left=1in, right=1in]{geometry}

\usepackage{microtype}
\usepackage{lineno}  
\usepackage{xspace} 
\usepackage{caption} 

\usepackage{graphicx}  
\usepackage{color}
\usepackage{colortbl}
\graphicspath{{./}} 

\usepackage{amsmath} 
\usepackage{amssymb}
\usepackage{amsfonts}
\usepackage{upgreek} 

\newcommand*\patchAmsMathEnvironmentForLineno[1]{%
\expandafter\let\csname old#1\expandafter\endcsname\csname #1\endcsname
\expandafter\let\csname oldend#1\expandafter\endcsname\csname
end#1\endcsname
 \renewenvironment{#1}%
   {\linenomath\csname old#1\endcsname}%
   {\csname oldend#1\endcsname\endlinenomath}%
}
\newcommand*\patchBothAmsMathEnvironmentsForLineno[1]{%
  \patchAmsMathEnvironmentForLineno{#1}%
  \patchAmsMathEnvironmentForLineno{#1*}%
}
\AtBeginDocument{%
\patchBothAmsMathEnvironmentsForLineno{equation}%
\patchBothAmsMathEnvironmentsForLineno{align}%
\patchBothAmsMathEnvironmentsForLineno{flalign}%
\patchBothAmsMathEnvironmentsForLineno{alignat}%
\patchBothAmsMathEnvironmentsForLineno{gather}%
\patchBothAmsMathEnvironmentsForLineno{multline}%
\patchBothAmsMathEnvironmentsForLineno{eqnarray}%
}

\usepackage{hyperxmp}

\usepackage[
            pdfauthor={\paperauthors},
            pdftitle={\paperasciititle},
            pdfkeywords={\paperkeywords},
            pdfcopyright={Copyright (C) \papercopyright},
            pdflicenseurl={\paperlicenceurl}]{hyperref}

\usepackage[colorinlistoftodos,textsize=scriptsize]{todonotes}

\usepackage[bottom,flushmargin,hang,multiple]{footmisc}

\usepackage[all]{hypcap} 

\input{lhcb-symbols-def} 

\usepackage{cite} 
\usepackage{mciteplus}

%% file: lhcb-symbols-def.tex
\usepackage{xspace} 
\usepackage{upgreek}

\def\lhcb   {\mbox{LHCb}\xspace}

\def\MagUp {\mbox{\em Mag\kern -0.05em Up}\xspace}

\ifthenelse{\boolean{uprightparticles}}%
{

 \def\Pmu         {\ensuremath{\upmu}\xspace}

 \def\Ppi         {\ensuremath{\uppi}\xspace}

 \def\Ppsi        {\ensuremath{\uppsi}\xspace}

 \def\PDelta      {\ensuremath{\Delta}\xspace}                 
 \def\PXi         {\ensuremath{\Xi}\xspace}                 
 \def\PLambda     {\ensuremath{\Lambda}\xspace}                 
 \def\PSigma      {\ensuremath{\Sigma}\xspace}                 
 \def\POmega      {\ensuremath{\Omega}\xspace}                 
 \def\PUpsilon    {\ensuremath{\Upsilon}\xspace}
 \let\oldPi\Pi
 \def\PPi         {\ensuremath{\oldPi}\xspace}

 \def\PB      {\ensuremath{\mathrm{B}}\xspace}                 
                  
 \def\PD      {\ensuremath{\mathrm{D}}\xspace}

 \def\PJ      {\ensuremath{\mathrm{J}}\xspace}                 
 \def\PK      {\ensuremath{\mathrm{K}}\xspace}                 
 \def\PL      {\ensuremath{\mathrm{L}}\xspace}

 \def\Pb      {\ensuremath{\mathrm{b}}\xspace}                 
 \def\Pc      {\ensuremath{\mathrm{c}}\xspace}                 
 \def\Pd      {\ensuremath{\mathrm{d}}\xspace}

 \def\Pi      {\ensuremath{\mathrm{i}}\xspace}

 \def\Pp      {\ensuremath{\mathrm{p}}\xspace}

 \def\Ps      {\ensuremath{\mathrm{s}}\xspace}                 
                  
 \def\Pu      {\ensuremath{\mathrm{u}}\xspace}

 \def\thebaroffset{0.0em}
}
{

 \def\Pmu         {\ensuremath{\mu}\xspace}

 \def\Ppi         {\ensuremath{\pi}\xspace}

 \def\Ppsi        {\ensuremath{\psi}\xspace}                 
                  
 \mathchardef\PDelta="7101
 \mathchardef\PXi="7104
 \mathchardef\PLambda="7103
 \mathchardef\PSigma="7106
 \mathchardef\POmega="710A
 \mathchardef\PUpsilon="7107
 \mathchardef\PPi="7105
                  
 \def\PB      {\ensuremath{B}\xspace}                 
                  
 \def\PD      {\ensuremath{D}\xspace}

 \def\PJ      {\ensuremath{J}\xspace}                 
 \def\PK      {\ensuremath{K}\xspace}                 
 \def\PL      {\ensuremath{L}\xspace}

 \def\Pb      {\ensuremath{b}\xspace}                 
 \def\Pc      {\ensuremath{c}\xspace}                 
 \def\Pd      {\ensuremath{d}\xspace}

 \def\Pi      {\ensuremath{i}\xspace}

 \def\Pp      {\ensuremath{p}\xspace}

 \def\Ps      {\ensuremath{s}\xspace}                 
                  
 \def\Pu      {\ensuremath{u}\xspace}

 \def\thebaroffset{0.18em}
}
\newcommand{\offsetoverline}[2][\thebaroffset]{\kern #1\overline{\kern -#1 #2}}%

\makeatletter
\ifcase \@ptsize \relax
  \newcommand{\miniscule}{\@setfontsize\miniscule{4}{5}}
\or
  \newcommand{\miniscule}{\@setfontsize\miniscule{5}{6}}
\or
  \newcommand{\miniscule}{\@setfontsize\miniscule{5}{6}}
\fi
\makeatother

\DeclareRobustCommand{\optbar}[1]{\shortstack{{\miniscule (\rule[.5ex]{1.25em}{.18mm})}
  \\ [-.7ex] $#1$}}

\def\mup        {{\ensuremath{\Pmu^+}}\xspace}
\def\mun        {{\ensuremath{\Pmu^-}}\xspace} 

\def\mumu       {{\ensuremath{\Pmu^+\Pmu^-}}\xspace}

\def\uquark    {{\ensuremath{\Pu}}\xspace}

\def\dquark    {{\ensuremath{\Pd}}\xspace}

\def\squark    {{\ensuremath{\Ps}}\xspace}

\def\cquark    {{\ensuremath{\Pc}}\xspace}

\def\bquark    {{\ensuremath{\Pb}}\xspace}

\def\pion   {{\ensuremath{\Ppi}}\xspace}

\def\pip    {{\ensuremath{\pion^+}}\xspace}
\def\pim    {{\ensuremath{\pion^-}}\xspace}

\def\kaon    {{\ensuremath{\PK}}\xspace}

\def\KorKbar {\kern \thebaroffset\optbar{\kern -\thebaroffset \PK}{}\xspace}

\def\Km      {{\ensuremath{\kaon^-}}\xspace}

\def\KS      {{\ensuremath{\kaon^0_{\mathrm{S}}}}\xspace}

\def\Kstar   {{\ensuremath{\kaon^*}}\xspace}

\def\D       {{\ensuremath{\PD}}\xspace}

\def\DorDbar {\kern \thebaroffset\optbar{\kern -\thebaroffset \PD}\xspace}
\def\Dz      {{\ensuremath{\D^0}}\xspace}

\def\Dp      {{\ensuremath{\D^+}}\xspace}
\def\Dm      {{\ensuremath{\D^-}}\xspace}

\def\DpDm    {\ensuremath{\Dp {\kern -0.16em \Dm}}\xspace}

\def\Dsm     {{\ensuremath{\D^-_\squark}}\xspace}

\def\B       {{\ensuremath{\PB}}\xspace}

\def\BorBbar {\kern \thebaroffset\optbar{\kern -\thebaroffset \PB}\xspace}

\def\Bd      {{\ensuremath{\B^0}}\xspace}

\def\BdorBdbar {\kern \thebaroffset\optbar{\kern -\thebaroffset \Bd}\xspace}
\def\Bu      {{\ensuremath{\B^+}}\xspace}
\def\Bub     {{\ensuremath{\B^-}}\xspace}
\def\Bp      {{\ensuremath{\Bu}}\xspace}
\def\Bm      {{\ensuremath{\Bub}}\xspace}

\def\Bs      {{\ensuremath{\B^0_\squark}}\xspace}

\def\BsorBsbar {\kern \thebaroffset\optbar{\kern -\thebaroffset \Bs}\xspace}

\def\jpsi     {{\ensuremath{{\PJ\mskip -3mu/\mskip -2mu\Ppsi}}}\xspace}

\def\Y#1S{\ensuremath{\PUpsilon{(#1S)}}\xspace}

\def\proton      {{\ensuremath{\Pp}}\xspace}
\def\antiproton  {{\ensuremath{\overline \proton}}\xspace}

\def\Lz          {{\ensuremath{\PLambda}}\xspace}
\def\Lbar        {{\ensuremath{\offsetoverline{\PLambda}}}\xspace}
\def\LorLbar     {\kern \thebaroffset\optbar{\kern -\thebaroffset \PLambda}\xspace}

\def\Xires       {{\ensuremath{\PXi}}\xspace}

\def\Lc          {{\ensuremath{\Lz^+_\cquark}}\xspace}

\def\Xicp        {{\ensuremath{\Xires^+_\cquark}}\xspace}

\def\Lb           {{\ensuremath{\Lz^0_\bquark}}\xspace}

\def\Xibm         {{\ensuremath{\Xires^-_\bquark}}\xspace}

\newcommand{\decay}[2]{\ensuremath{#1\!\to #2}\xspace} 

\def\to                 {\ensuremath{\rightarrow}\xspace}

\def\CP                {{\ensuremath{C\!P}}\xspace}

\def\AT#1     {\ensuremath{A_{\mathrm{T}}^{#1}}\xspace}           

\def\C#1      {\ensuremath{\mathcal{C}_{#1}}\xspace}                       
\def\Cp#1     {\ensuremath{\mathcal{C}_{#1}^{'}}\xspace}                    
\def\Ceff#1   {\ensuremath{\mathcal{C}_{#1}^{\mathrm{(eff)}}}\xspace}        
\def\Cpeff#1  {\ensuremath{\mathcal{C}_{#1}^{'\mathrm{(eff)}}}\xspace}       
\def\Ope#1    {\ensuremath{\mathcal{O}_{#1}}\xspace}                       
\def\Opep#1   {\ensuremath{\mathcal{O}_{#1}^{'}}\xspace}                    

\newcommand{\braket}[2]{\ensuremath{\langle #1|#2\rangle}} 

\newcommand{\aunit}[1]{\ensuremath{\text{\,#1}}}       

\newcommand{\tev}{\aunit{Te\kern -0.1em V}\xspace}
\newcommand{\gev}{\aunit{Ge\kern -0.1em V}\xspace}
\newcommand{\mev}{\aunit{Me\kern -0.1em V}\xspace}
\newcommand{\kev}{\aunit{ke\kern -0.1em V}\xspace}
\newcommand{\ev}{\aunit{e\kern -0.1em V}\xspace}
 
\newcommand{\mevc}{\ensuremath{\aunit{Me\kern -0.1em V\!/}c}\xspace}
\newcommand{\gevc}{\ensuremath{\aunit{Ge\kern -0.1em V\!/}c}\xspace}
\newcommand{\mevcc}{\ensuremath{\aunit{Me\kern -0.1em V\!/}c^2}\xspace}
\newcommand{\gevcc}{\ensuremath{\aunit{Ge\kern -0.1em V\!/}c^2}\xspace}

\def\fb   {\ensuremath{\aunit{fb}}\xspace}
\def\invfb   {\ensuremath{\fb^{-1}}\xspace}

\newcommand{\chisq}{\ensuremath{\chi^2}\xspace}

\def\deriv {\ensuremath{\mathrm{d}}}

\def\gsim{{~\raise.15em\hbox{$>$}\kern-.85em
          \lower.35em\hbox{$\sim$}~}\xspace}
\def\lsim{{~\raise.15em\hbox{$<$}\kern-.85em
          \lower.35em\hbox{$\sim$}~}\xspace}

\def\sPlot{\mbox{\em sPlot}\xspace}

\def\pt         {\ensuremath{p_{\mathrm{T}}}\xspace}

\def\ptot       {\ensuremath{p}\xspace}

\def\evtgen     {\mbox{\textsc{EvtGen}}\xspace}

\def\geant      {\mbox{\textsc{Geant4}}\xspace}

\def\photos     {\mbox{\textsc{Photos}}\xspace}

\def\pythia     {\mbox{\textsc{Pythia}}\xspace}

\def\tell1  {TELL1\xspace}
\def\ukl1   {UKL1\xspace}

\newcommand{\eg}{\mbox{\itshape e.g.}\xspace}
\newcommand{\ie}{\mbox{\itshape i.e.}\xspace}

\newcommand{\lhcborcid}[1]{\href{https://orcid.org/#1}{\hspace*{0.1em}\raisebox{-0.45ex}{\includegraphics[width=1em]{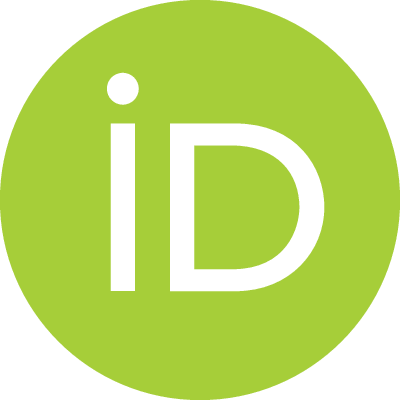}}}}


%% file: title-LHCb-PAPER.tex

\begin{titlepage}
\pagenumbering{roman}

\vspace*{-1.5cm}
\centerline{\large EUROPEAN ORGANIZATION FOR NUCLEAR RESEARCH (CERN)}
\vspace*{1.5cm}
\noindent
\begin{tabular*}{\linewidth}{lc@{\extracolsep{\fill}}r@{\extracolsep{0pt}}}
\ifthenelse{\boolean{pdflatex}}
{\vspace*{-1.5cm}\mbox{\!\!\!\includegraphics[width=.14\textwidth]{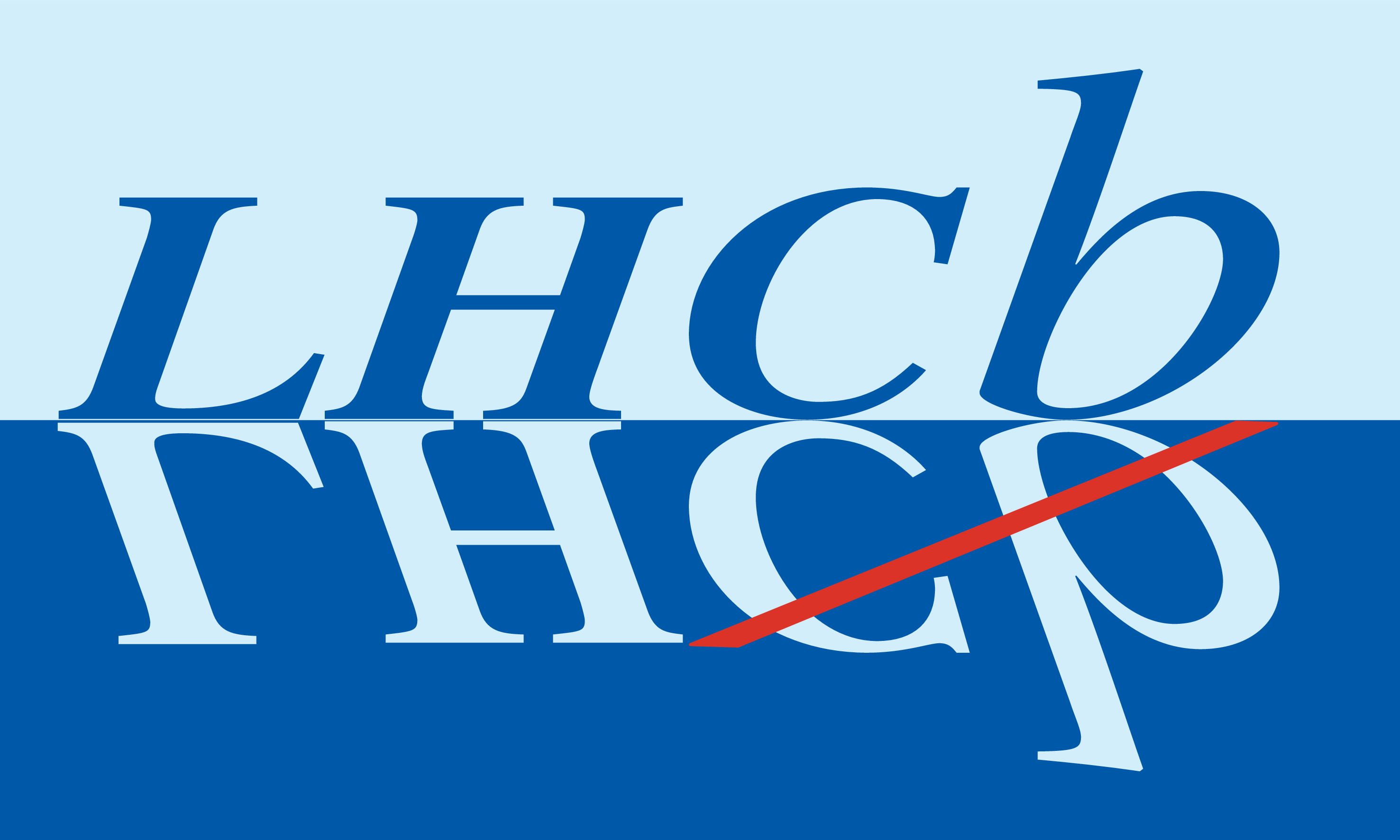}} & &}%
{\vspace*{-1.2cm}\mbox{\!\!\!\includegraphics[width=.12\textwidth]{lhcb-logo.eps}} & &}%
\\
 & & CERN-EP-2022-198 \\  
 & & LHCb-PAPER-2022-031 \\  
 & & June 12, 2024 \\ 
\\
\end{tabular*}

\vspace*{4.0cm}

{\normalfont\bfseries\boldmath\huge
\begin{center}
  \papertitle 
\end{center}
}

\vspace*{2.0cm}

\begin{center}
\paperauthors\footnote{Authors are listed at the end of this paper.}
\end{center}

\vspace{\fill}

\begin{abstract}
  \noindent
An amplitude analysis of $\decay{\Bm}{\jpsi\Lz\antiproton}$ decays is performed using 4400 signal candidates selected on a data sample of $\proton\proton$ collisions recorded at center-of-mass energies of 7, 8 and 13 \tev with the \lhcb detector, corresponding to an integrated luminosity of 9 \invfb. 
A narrow resonance in the \jpsi\Lz system, consistent with a pentaquark candidate with strangeness, is observed with high significance. The mass and the width of this new state are measured to be $4338.2\pm 0.7\pm 0.4\mev$ and ${7.0\pm1.2\pm1.3\mev}$, where the first uncertainty is statistical and the second systematic. The spin is determined to be 
$1/2$ and negative parity is preferred. Due to the small $Q$-value of the reaction, the  most precise single measurement of the \Bm mass to date, $5279.44\pm0.05\pm0.07\mev$, is obtained.
\end{abstract}

\vspace*{2.0cm}

\begin{center}
  Published in
  Phys.~Rev.~Lett. 131 (2023) 031901
\end{center}

\vspace{\fill}

{\footnotesize 
\centerline{\copyright~\papercopyright. \href{\paperlicenceurl}{\paperlicence}.}}
\vspace*{2mm}

\end{titlepage}


\newpage
\setcounter{page}{2}
\mbox{~}
%
%
%
%

%% file: body.tex

The discovery of pentaquark candidates in the \jpsi\proton system at \lhcb~\cite{LHCb-PAPER-2015-029,LHCb-PAPER-2019-014} opened a new field of investigation in baryon spectroscopy.
Such resonant structures with valence quark content (the  exotic hadron naming scheme defined in Ref.~\cite{Gershon:2814506} is used throughout this Letter) ${P_{\psi}^{N}}^+$ = \cquark\!\!$\overline{\cquark}$\uquark\uquark\dquark have been observed only in the $\Lb\to\jpsi\proton\Km$ decay to date.
Recently, evidence for a new candidate was found in the $\Bs\to\jpsi\proton\antiproton$ decay~\cite{LHCb-PAPER-2018-046, LHCb-PAPER-2021-018}, 
and evidence for a ${P_{\psi s}^{\Lz}}^{0}$ = \cquark\!\!$\overline{\cquark}$\uquark\dquark\squark pentaquark candidate with strangeness was found in the \jpsi\Lz system in the $\decay{\Xibm}{\jpsi\Lz\Km}$ decay~\cite{LHCb-PAPER-2020-039} (charge conjugation is implied throughout this Letter).  

Pentaquarks are predicted within the quark model to have a minimal quark content of three quarks plus a quark-antiquark pair. 
Experimentally, the pentaquark candidates are found close to threshold for the production of ordinary baryon-meson states, \ie $\varSigma_c^+\overline{D}^0$ and $\varSigma_c^+\overline{D}^{* 0}$ for the observed ${P_{\psi}^{N}}^+$ states~\cite{LHCb-PAPER-2015-029,LHCb-PAPER-2019-014}, and ${\varXi}_c^0\overline{D}^{* 0}$ for the ${P_{\psi s}^{\Lz}}^0$ state~\cite{LHCb-PAPER-2020-039}.
Various interpretations of these states have been proposed, including tightly bound pentaquark states~\cite{Esposito:2016noz, Richard:2016eis}, loosely bound baryon-meson molecular states (Refs.\hspace{1sp}\cite{Dong:2021juy, Guo:2017jvc} and references therein), and rescattering effects~\cite{Guo:2019twa}. Hidden-charm pentaquarks with strangeness were predicted in~\cite{Xiao:2019gjd,Wang:2019nvm} as hadronic molecules, and in Ref.~\cite{Ali:2019clg} as compact states. However, their nature is still largely unknown and further investigation is needed~\cite{Olsen:2017bmm}.

The \decay{\Bm}{\jpsi\Lz\antiproton} decay offers the unique opportunity to simultaneously search for ${\overline{P}_{\psi}^{N}}^-$ and  ${P_{\psi s}^{\Lz}}^0$ pentaquark candidates in the \jpsi\antiproton and \jpsi\Lz systems, respectively. In particular, the phase space available in the decay allows searches for pentaquark candidates located close to different baryon-meson thresholds, such as \Lc\Dz for ${P_{\psi}^{N}}^+$, and  \Lc\Dsm, \Xicp\Dm for ${P_{\psi s}^{\Lz}}^0$ candidates. Neither the ${P_{\psi s}^{\Lz}(4459)}^{0}$ state, found in the $\Xibm\to \jpsi\Lz\Km$ decay~\cite{LHCb-PAPER-2020-039}, nor the ${P_{\psi}^{N}(4337)}^+$ state, found in the $\Bs\to\jpsi\proton\antiproton$ decays~\cite{LHCb-PAPER-2021-018}, is accessible with the present analysis since they are outside of the available phase space.

The small $Q$-value of the decay, approximately (natural units
with $\hslash = c = 1$ are used throughout this Letter) $128\mev$, 
provides excellent mass resolution, allowing searches for narrow resonant structures.
In addition, efficient reconstruction of low momentum tracks can improve sensitivity to resonance structures near threshold. This decay was previously studied by the CMS collaboration using a sample of $450\pm20$ signal candidates and the invariant mass distributions of the \jpsi\Lz, \jpsi\antiproton, \Lz\antiproton systems were found to be inconsistent with the pure phase-space hypothesis~\cite{CMS:2019kbn}.  
In this Letter, an amplitude analysis of the \decay{\Bm}{\jpsi\Lz\antiproton} decay is performed using signal candidates selected on a data sample of $\proton\proton$ collisions at centre-of-mass energies of 7\tev and 8\tev (Run 1),  and 13\tev (Run 2), recorded between 2011 and 2018 by the \lhcb detector, corresponding to an integrated luminosity of 9 \invfb. In the following, the first observation of a ${P_{\psi s}^{\Lz}}^0$ pentaquark candidate with strangeness in the $\jpsi\Lz$ system is reported, which is different from the ${P_{\psi s}^{\Lz}(4459)}^0$ state found in the $\decay{\Xibm}{\jpsi\Lz\Km}$ decay~\cite{LHCb-PAPER-2020-039}.


The \lhcb detector is a single-arm forward
spectrometer covering the pseudorapidity range $2<\eta <5$, described in detail in Refs.~\cite{LHCb-DP-2014-002, LHCb-DP-2014-001, LHCb-DP-2013-003, LHCb-DP-2012-002}.
The online event selection is performed by a trigger~\cite{LHCb-DP-2012-004}, comprising a hardware stage based on information from the muon system which selects $\jpsi\to \mup\mun$ decays, followed by a software stage that applies a full event reconstruction. The software trigger relies on identifying \jpsi decays into muon pairs consistent with originating from a \B-meson decay vertex detached from the primary \proton\proton collision point.

Samples of simulated events are used to study the properties of the signal mode decay ${\Bm\to\jpsi\Lz(\to\proton\pim)\antiproton}$  and the control-mode  decay ${\Bm\to\jpsi\Kstar(892)^-[\to\KS(\to\pip\pim)\pim]}$. The latter are used to calibrate the distributions of simulated \Bm decays with data. 

The \proton\proton collisions are generated using
\pythia~\cite{Sjostrand:2007gs} with a specific \lhcb
configuration~\cite{LHCb-PROC-2010-056}. Decays of hadronic particles and interactions with the detector material are described by \evtgen~\cite{Lange:2001uf}, using \photos~\cite{davidson2015photos}, and by the \geant toolkit~\cite{Allison:2006ve, Agostinelli:2002hh, LHCb-PROC-2011-006}, respectively. The signal and the control-mode decays are generated from a uniform phase-space distribution. 


Signal \Bm candidates are formed from combinations of \jpsi, \Lz and \antiproton candidates originating from a common decay vertex. The \jpsi candidates are formed from pairs of oppositely charged tracks identified as muons and originating from a decay vertex significantly displaced from the associated \proton\proton primary vertex (PV). 
The associated PV for a given particle is the PV with the smallest impact parameter $\chi^2_{\rm IP}$, defined as the difference in the vertex-fit \chisq of a given PV reconstructed with and without the particle under consideration.
The ${\Lz\to\proton\pim}$ candidates are formed from pairs of oppositely charged tracks and selected in two different categories according to the \Lz decay position: i) the ``long'' category for early decays that allow the proton and pion candidates to be reconstructed in the vertex detector; ii) the ``downstream'' category for \Lz baryons that decay outside the vertex detector and are reconstructed in the tracking stations only. The long candidates have  better mass, momentum and vertex resolution than downstream candidates. The \antiproton candidate is a charged track identified as an antiproton. 

A kinematic fit~\cite{Hulsbergen:2005pu} to the \Bm candidate is performed with the dimuon and the \proton\pim masses constrained to the known \jpsi and \Lz masses, respectively~\cite{PDG2022}. 
 Simulated events are weighted such that the distributions of transverse momentum ($\pt$) and number of tracks per event for \Bm candidates match the \decay{\Bm}{\jpsi\Kstar(892)^-} control-mode distributions in data. 
In simulation, the particle identification (PID) variables for each charged track are resampled as a function of their \ptot, $\pt$ and the number of tracks in the event using $\decay{\Lc}{\proton\Km\pip}$  
 and $D^{\ast +}\to D^0(\to K^-\pip)\pi^+$ calibration samples from data~\cite{LHCb-DP-2018-001}. 
 
 The final stage of the selection uses multivariate techniques trained with simulation and data. Separate boosted-decision-tree (BDT,~\cite{Breiman}) classifiers are employed for the four combinations of two data-taking periods (Run~1 and Run~2) and two signal categories, using long and downstream reconstructed \Lz candidates. Each BDT is trained on simulated signal decays and data sidebands, with the $m(\jpsi \Lz\antiproton)$ invariant mass in the range $[5320, 5360]\mev$.  
 The variables used as input to the BDT are the $\pt$, the decay length significance, the angle between the momentum and the flight direction and the $\chi^2_{\rm IP}$ variable of the \Bm candidate; the $\chi^2$ probability from the kinematic fit of the candidate; the sum of the $\chi^2_{\rm IP}$ of the daughter particles; the angle between the momentum and the flight direction, the $\chi^2$ of the flight distance (only for long category candidates), the $\chi^2_{\rm IP}$ variables of the \Lz candidate, and the hadron PID for the \antiproton candidate from the ring-imaging Cherenkov detectors.
 
 The BDT output selection criterion is chosen as in Ref.~\cite{LHCb-PAPER-2021-018} by maximising the figure of merit $\mathcal{S}^2/(\mathcal{S}+\mathcal{B})^{3/2}$ to obtain both high signal purity and significance, where $\mathcal{S}$ and $\mathcal{B}$ are the signal and background yield in a region of $\pm5.3\mev$ around the  known \Bm mass. To avoid a possible bias due to fluctuations of the signal yield, $\mathcal{S}$  is determined from a fit to the \jpsi \Lz \antiproton invariant-mass distribution in data after applying a loose BDT selection, multiplied by the efficiency of the BDT output requirement obtained from simulation.  Similarly, $\mathcal{B}$ is extracted from a fit to sideband data.

\begin{figure}[t]
  \centering
  \includegraphics[width=0.5\textwidth]{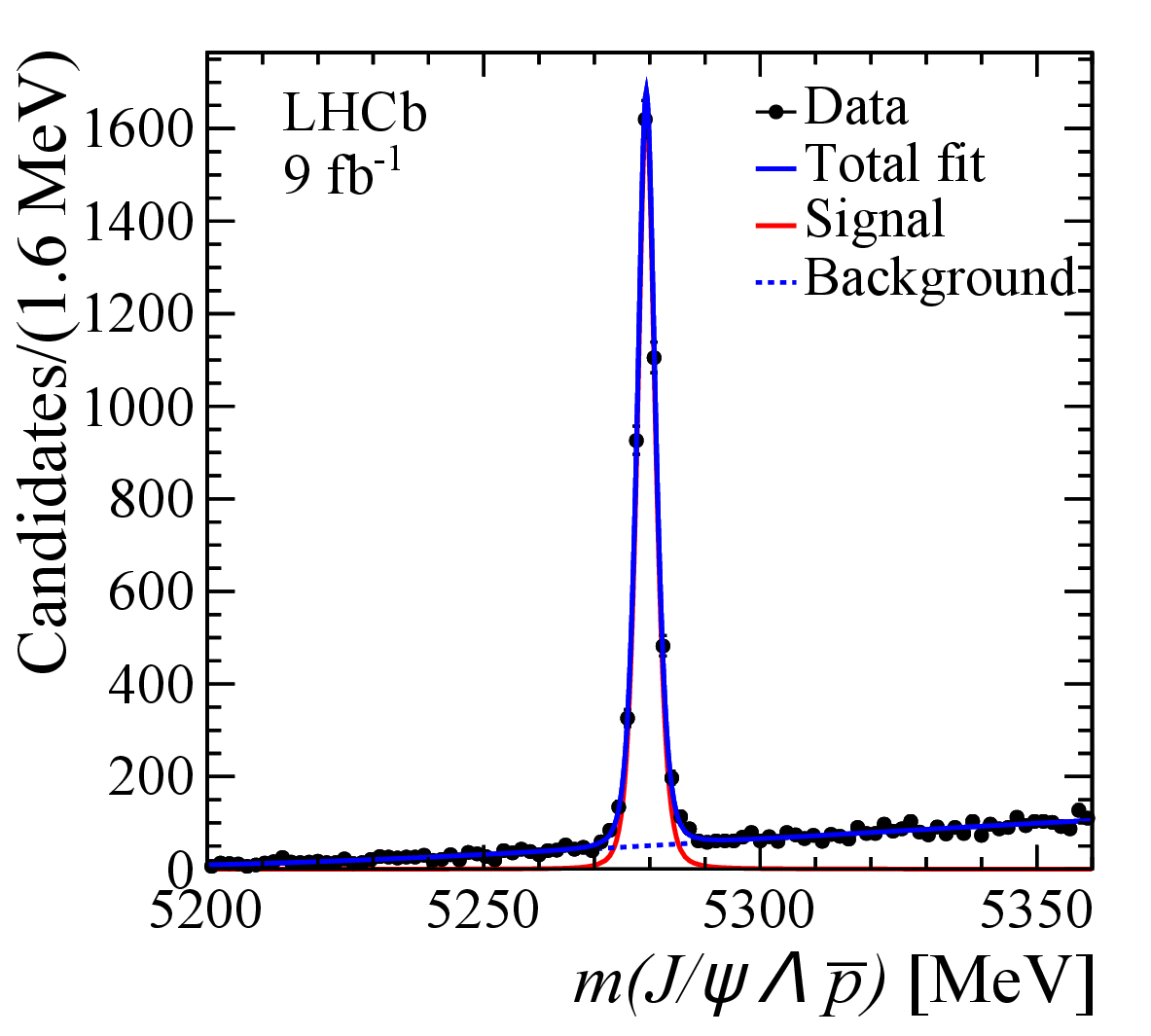}
  \caption{Invariant mass distribution of the \jpsi \Lz \antiproton candidates. The data are overlaid with the results of the fit.}
  \label{fig:fit_mass}
\end{figure}

For candidates passing all selection criteria, a maximum-likelihood fit is performed to the $m(\jpsi \Lz \antiproton)$ distribution shown in Fig.~\ref{fig:fit_mass}, resulting in a signal yield of $4620\pm70$. For the amplitude analysis about 4400 signal candidates are retained, 
with a purity of $93.0\%$ in the signal region of $\pm 2.5 \sigma$ around the mass peak, where $\sigma\approx 2.1\mev$ is the mass resolution. The signal distribution is modelled by the sum of a Johnson function~\cite{JohnsonSU} and two Crystal Ball~\cite{Skwarnicki:1986xj} functions sharing the same mean
and width
parameters determined from the fit. The tail parameters and fractions of each signal component are fixed to values obtained from a fit to simulated events. The background contribution is mainly due to random combinations of charged particles in the event and is described by a third-order Chebyshev polynomial.

\begin{figure}[t]
  \centering
  
  \includegraphics[width=0.5\textwidth]{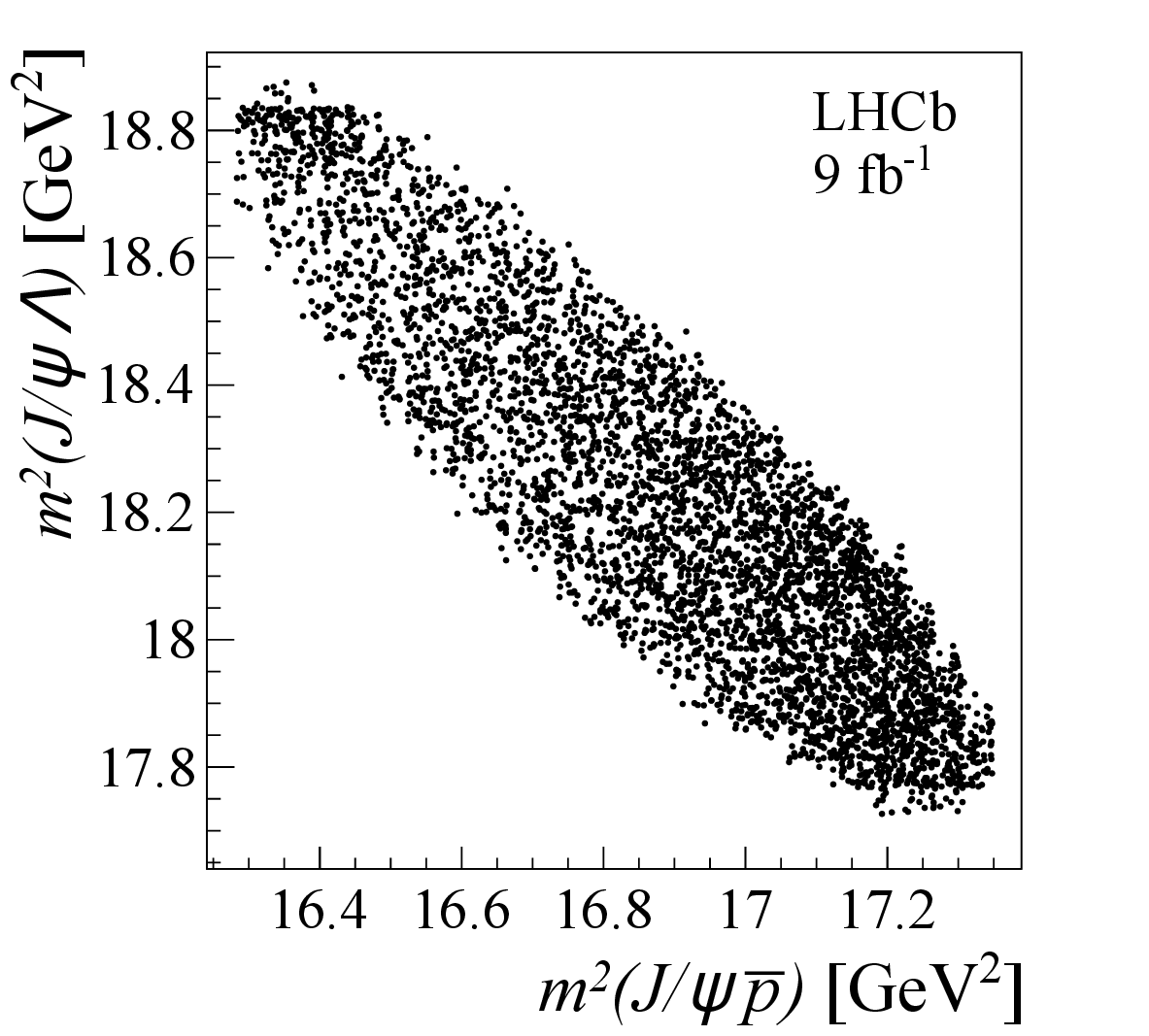}
  \caption{Dalitz distribution for \Bm candidates in the signal region.}
   
  \label{fig:Dalitz}
\end{figure}

The Dalitz distribution of the reconstructed \Bm candidates in the signal region is shown in Fig.~\ref{fig:Dalitz}, where a horizontal band in the region around $18.8\gev^2$ in the $m^2(\jpsi\Lz)$ distribution is present. Some structure in the high $m^2(\jpsi\antiproton)$ spectrum is also present. This Letter investigates the nature of these enhancements.

An amplitude analysis of the \Bm candidates in the signal region is performed using a phenomenological model based on the interference of two-body resonances in the three decay chains, $\jpsi K^{*-}(\to \Lz \antiproton)$, $\Lz \overline{P}_{\psi}^{N-}(\to\jpsi\antiproton)$, and $\antiproton {P_{\psi s}^{\Lz}}^0(\to \jpsi\Lz)$, labelled as the $K^{*-}$, ${\overline{P}_{\psi}^{N}}^-$ and  ${P_{\psi s}^{\Lz}}^0$ chains, respectively.
The angular information of the subsequent \decay{\jpsi}{\mumu} and \decay{\Lz}{\proton\pim} decays are taken into account in all cases. The decay amplitudes are based on helicity formalism~\cite{Chung:186421} with \CP symmetry enforced, and follow the prescriptions in Ref.\cite{DPdeco} for the spin alignment of the different decay chains. Details about the decay amplitude definition are given in the Supplemental material~\cite{SupplementalMaterial}.

The decay amplitudes are defined as a function of the six-dimensional phase space of the \Bm decay, $(m_{\Lz\antiproton},\vec{\Omega})$ described by the combined invariant mass $m_{\Lz\antiproton}$ of the \antiproton and \Lz pairs, and by five angular variables indicated as $\vec{\Omega}$: the cosine of the helicity angle, $\cos\theta_{\Kstar}$ $(\cos\theta_{\jpsi})$ of the \Lz $(\mun)$ in the \Lz\antiproton  (\jpsi) rest frame, the azimuthal angle $\phi_\proton$ $(\phi_\mun)$ of the $\proton$ ($\mun$) in the rest frame of the \Lz (\jpsi), and the cosine of the helicity angle $\cos\theta_{\Lz}$, of the \proton in the rest frame of the \Lz. 
The amplitude fit to determine the model parameters $\vec{\omega}$, \ie, the couplings, the masses, the widths, and lineshape parameters of different contributions, is performed by minimising the negative log-likelihood function,
\begin{align}
\label{eq:Likelihood}
-2\log{{\cal L}}(\vec{\omega})=-2\sum_{i}\log{\left[(1-\beta){\cal P}_{\rm{sig}} (m_{\Lz\antiproton,i},\vec{\Omega_i}|\vec{\omega})+\beta{\cal P_{\rm{bkg}}}(m_{\Lz\antiproton,i},\vec{\Omega}_i)\right]},   
\end{align}
where ${\cal P}_{\rm{sig}}$ $({\cal P}_{\rm{bkg}})$ is the probability density function (PDF) for the signal (background) component of the $i$th event, and $\beta=0.07\pm 0.01$ is the fraction of background candidates in the signal region. The signal PDF is proportional to the squared decay amplitude $|{\cal M}(m_{\Lz\antiproton},\vec{\Omega}|\vec{\omega})|^2$, and accounts for the phase-space element $\Phi(m_{\Lz\antiproton})$ and the reconstruction efficiency $\epsilon(m_{\Lz\antiproton},\vec{\Omega})$,
\begin{equation}
\label{eq:Psig}    
{\cal P}_{\rm{sig}}(m_{\Lz\antiproton},\vec{\Omega}|\vec{\omega})=\frac{|{\cal M}(m_{\Lz\antiproton},\vec{\Omega}|\vec{\omega})|^2 \Phi(m_{\Lz\antiproton}) \epsilon(m_{\Lz\antiproton},\vec{\Omega})}{I(\vec{\omega})}.
\end{equation}
The denominator $I(\vec{\omega})$ normalizes the probability. The background PDF ${\cal P}_{\rm bkg}$ is parameterized according to a six-dimensional phase-space function based on Legendre polynomials, whose coefficients are determined from the $m(\jpsi\Lz\antiproton)$ region ${[5200, 5250] \cup [5340, 5350] \mev}$. Similarly, the reconstruction efficiency is parameterized using Legendre polynomials with coefficients determined using simulated phase-space signal decays. 

No well-established resonances are expected to decay into the $\jpsi\Lz$ and $\jpsi\antiproton$ final states. However, excited $K^{-}$ resonances decaying outside of the phase space of the ${\Bm \to \jpsi \Lz \antiproton}$ decay can contribute to the $\Lz\antiproton$ channel~\cite{CMS:2019kbn}. 
A fit including only NR contributions and $K^*_4(2045)^-$, $K_2(2250)^-$, and $K_3(2320)^-$ resonant amplitudes does not reproduce the data distribution. A $\chisq/{\rm n.d.f.}$ of $123.2/46$ is obtained, where the $\chisq$ is calculated as the largest value over the six one-dimensional fit projections and the number of degrees of freedom (${\rm n.d.f.}$) is extracted from pseudoexperiments by fitting the tail of the $\chisq_{\rm max}$ distribution. 
The simplest and most effective amplitude model used to fit the data, indicated as the nominal model in the following, comprises a narrow $\jpsi\Lz$ structure with spin-parity $J^P = 1/2^-$, whose mass and width are extracted from the amplitude fit, and two nonresonant (NR) contributions, one with $J^P=1^-$ for the $\Lz\antiproton$ system and a second one with $J^P=1/2^-$ for the $\jpsi\antiproton$, referred to as NR($\Lz\antiproton$) and NR($\jpsi\antiproton$), respectively. The $\jpsi\Lz$ resonance is modelled with a relativistic Breit--Wigner function as discussed in the Supplemental material~\cite{SupplementalMaterial}.

The couplings are defined in the $LS$ basis, both for the $\Bm\to X R$ process, and for the $R\to YZ$ process, where $X$, $Y$, and $Z$ are the final state particles, and $R=K^{*-},{\overline{P}_{\psi}^{N}}^-$, and ${P_{\psi s}^{\Lz}}^0$ is the decay chain under consideration. 
Here, $L$ indicates the decay orbital angular momentum, and $S$ is the sum of the spins of the decay products. In the nominal model, $L=0$ is used for the production and decay of the narrow $\jpsi\Lz$ resonance, while $L=0,1,2$ and $L=0,2$  are used in the NR($\Lz\antiproton$) system for the production and decay, respectively, and  $L=0$ and $L=1$ in the NR($\jpsi\antiproton$) system. 
Because of the small $Q$-value of the decay, higher values of the orbital momentum are suppressed. Fixing the lowest orbital momentum couplings for the NR($\jpsi\antiproton$) as the normalization choice reduces the number of free parameters to 16: the mass, the width, and the complex coupling of the ${P_{\psi s}^{\Lz}}^{0}$ resonant contribution, four complex couplings for the NR($\Lz\antiproton$) contribution, and a complex coupling and two parameters for the second-order polynominal parameterization of the lineshape for the NR($\jpsi\antiproton$) contribution.  

\begin{figure}[t]
\centering
\small
\includegraphics[width=0.49\linewidth]{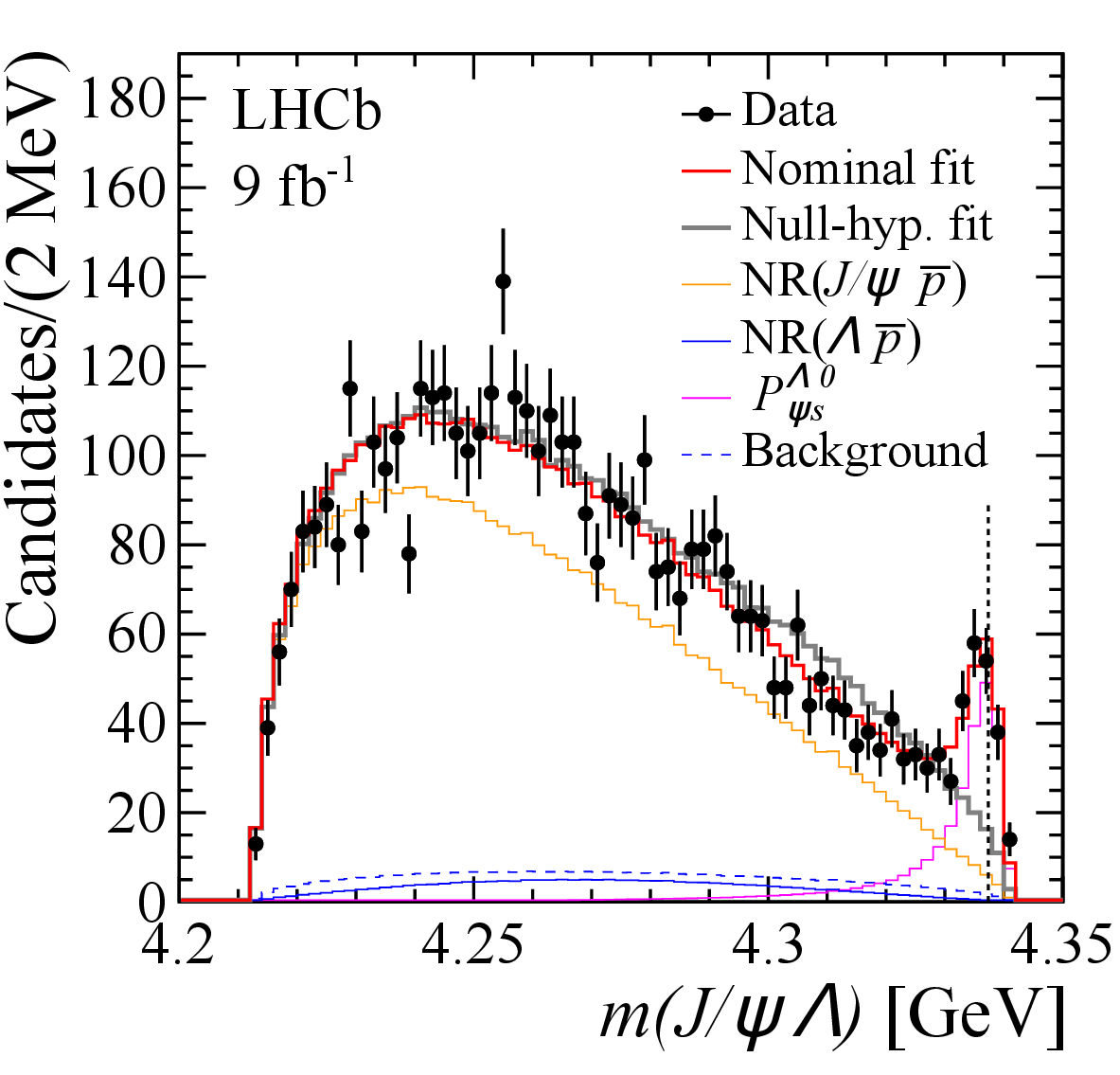}
\includegraphics[width=0.49\linewidth]{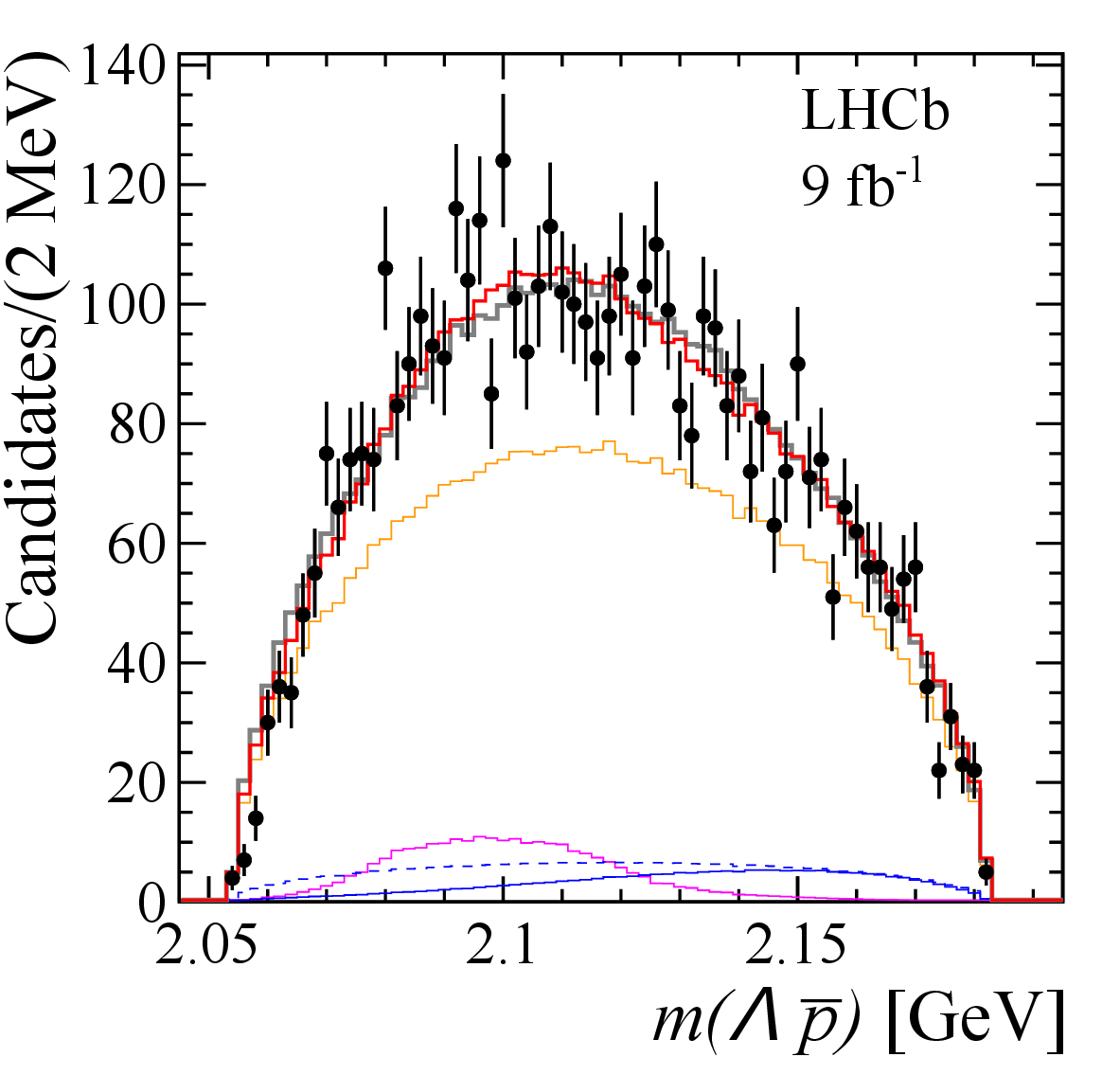} \\
\includegraphics[width=0.49\linewidth]{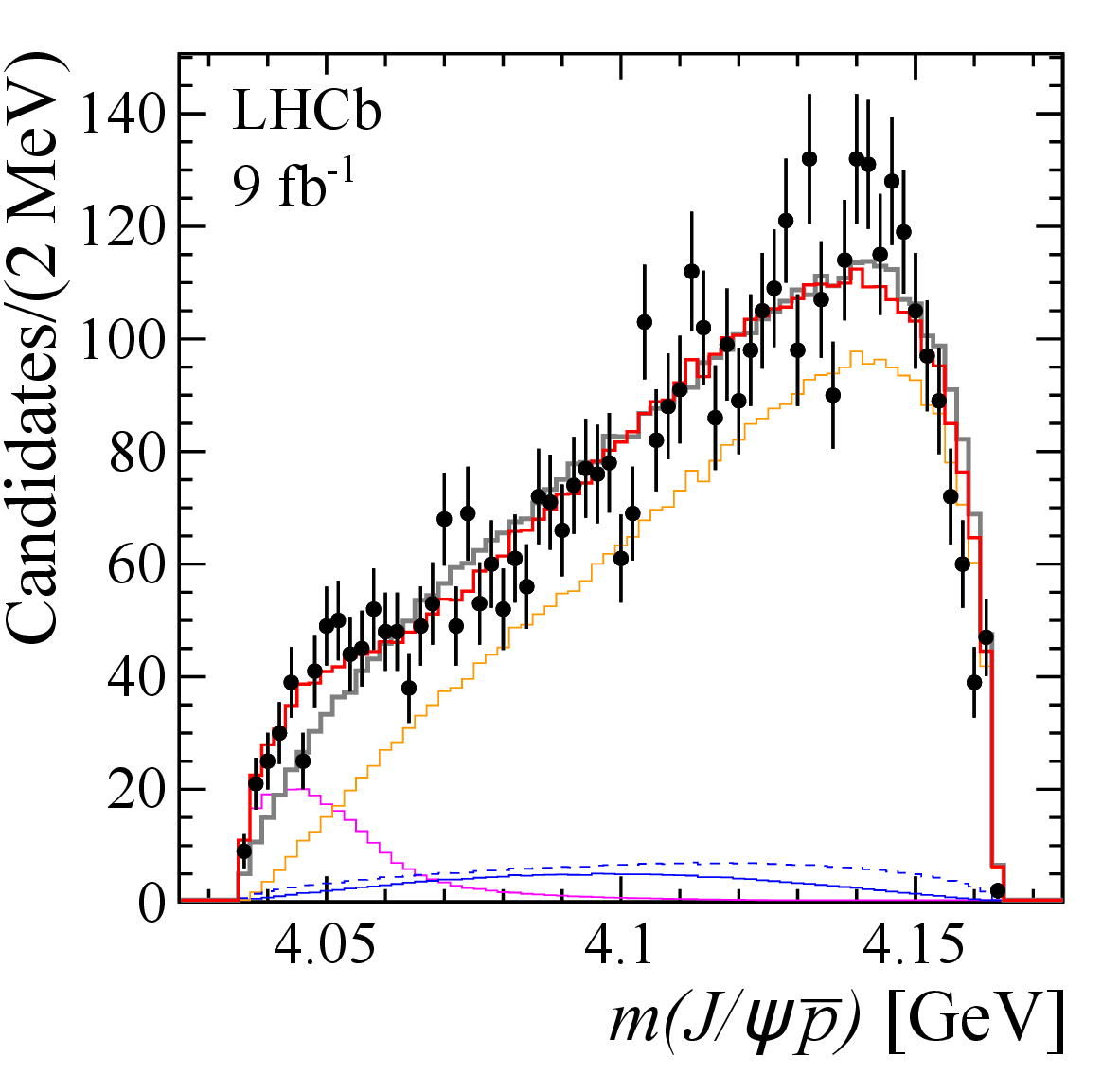}
\includegraphics[width=0.49\linewidth]{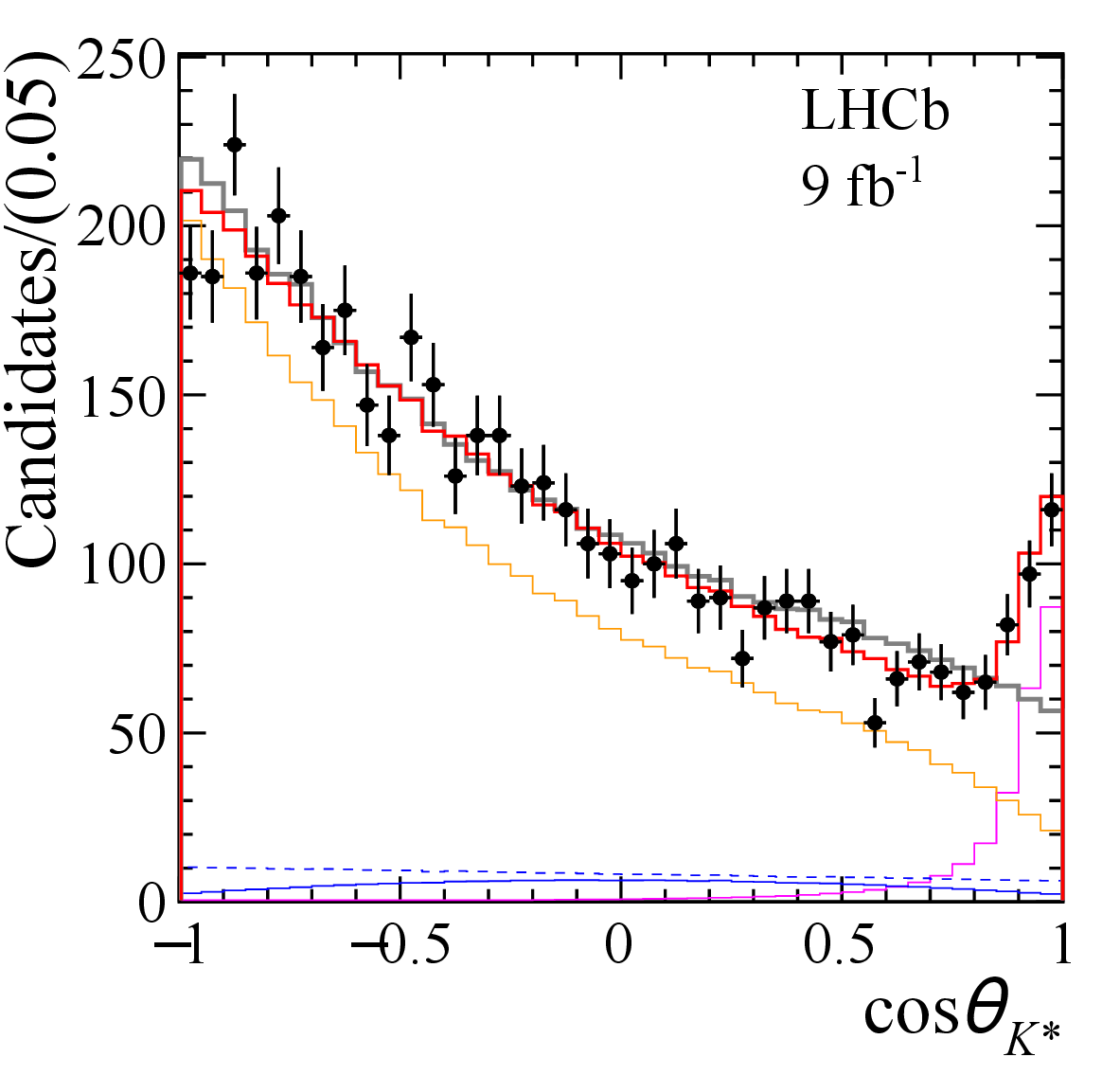}
  \caption{\small\label{fig:FullAmp_fit} Distributions of invariant mass and  $\cos\theta_{K^*}$. 
  Fit results to data using the nominal model are superimposed. The  null-hypothesis model fit results are also shown in grey. The $\varXi_c^+D^-$ baryon-meson threshold at 4.337 \gev is indicated with a vertical dashed line in the $m(\jpsi\Lz)$ invariant mass distribution.} 
\end{figure}

A null-hypothesis model is used to test the significance of the ${P_{\psi s}^{\Lz}}^0$ state, which comprises only two NR contributions. 
The fit results for the nominal and the null-hypothesis model are shown in Fig.~\ref{fig:FullAmp_fit}. The null-hypothesis model does not describe the data, with a corresponding $\chisq_{\rm max}/{\rm n.d.f.}=120.8/47$. 
Using the nominal model, a good fit to data was obtained with a $\chisq_{\rm max}/{\rm n.d.f.}=55.3/51$ and a $p$-value of $0.51$ computed by counting the number of pseudoexperiments above the value of $\chi^2_{\rm max}$ observed in data.

A new narrow $\jpsi\Lz$ structure is observed with high significance in the nominal fit to data.  Using Wilks's theorem, a statistical significance exceeding $15\sigma$ is estimated from the value of $-2\Delta\log{\cal L}=243$ of the null-hypothesis model with respect to the nominal model. The mass and width of the new pentaquark candidate are measured to be $M_{P_{\psi s}^{\Lz}} = 4338.2\pm 0.7 \mev$ and $\Gamma_{P_{\psi s}^{\Lz}} = 7.0\pm 1.2 \mev$, respectively, where the uncertainties are statistical only. 
This represents the first observation of a strange pentaquark candidate with minimal quark content \cquark\!\!$\overline{\cquark}$\uquark\dquark\squark.  

Alternative models are considered for systematic studies. 
To assess the contribution of a ${\overline{P}_{\psi}^{N}}^-$ pentaquark candidate, a relativistic Breit--Wigner function is used for the $m(\jpsi \antiproton)$ lineshape instead of a 2nd order polynomial function. The value of $-2\Delta \log {\cal L} = 80$ obtained with respect to the nominal fit indicates that the NR($\jpsi \antiproton$) contribution is preferred over the hypothesis of a ${\overline{P}_{\psi}^{N}}^-$ candidate, while consistent results for the ${P_{\psi s}^{\Lz}(4338)}^0$ state parameters are obtained. The contribution of a second narrow ${P_{\psi s}^{\Lz}}^0$ resonance is added to the nominal model to parametrize the $m(\jpsi\Lz)$ distribution close to the $\Lc\Dsm$ threshold at $4255 \mev$, and found to not be  statistically significant. 
Using the ${\rm CL}_s$ method~\cite{Read:2002hq}, an upper limit on
the ${P_{\psi s}^{\Lz}(4255)}^0$ fit fraction is set to 8.7\% 
at a 95\% confidence level.
To determine the $J^P$ assignments, all 16 combinations of $J^P=1/2^\pm,3/2^\pm$ are studied for the ${P_{\psi s}^{\Lz}(4338)}^0$ and NR($\jpsi\antiproton$) spin-parity hypotheses, and those with $-2\Delta \log {\cal L}>9$ with respect to the nominal fit are discarded. For the ${P_{\psi s}^{\Lz}(4338)}^0$ state, the $J^P=3/2^{\pm}$ hypotheses are discarded, the $J^P=1/2^-$ assignment is preferred, while the $J^P=1/2^+$ is excluded at a 90\% confidence level using the ${\rm CL}_s$ method~\cite{Read:2002hq}.    
%

Systematic uncertainties are evaluated on the mass and the width of the new pentaquark candidate, and on the fit fractions of ${P_{\psi s}^{\Lz}(4338)}^0$,  NR($\jpsi\antiproton$), and NR$(\jpsi\Lz)$ contributions. The uncertainties are summarised in Table~\ref{tab:syst} and are summed in quadrature for the total contribution. For each systematic uncertainty, an ensemble of 1000 pseudoexperiments, generated according to the nominal model with the same statistics as in data is fitted with an alternative configuration that is representative of the systematic effect. The uncertainty on each parameter is determined as the mean value of the difference between the fit results of the nominal and the alternative models.
The main contributions are related to the  model for the decay amplitude, the bias of the fitting procedure, and the uncertainty on the reconstruction efficiency $\epsilon(m_{\antiproton\Lz},\vec{\Omega})$.
 For the amplitude model, the nominal value of the hadron radius for the Blatt--Weisskopf coefficients~\cite{Blatt:1952ije} is assumed to be $3\gev^{-1}$ and varied to 1 and $5\gev^{-1}$, taking the largest effect as a systematic uncertainty. Additional $LS$ couplings are considered with respect to the nominal model, in particular, the $L,S=1,1$ ($L,S=2,3/2$) coupling for the production (decay) of ${P_{\psi s}^{\Lz}(4338)}^0$ contribution, and the $L,S=1,1$ coupling for the NR$(\jpsi\antiproton)$ contribution. A relativistic Breit--Wigner function is used instead of the 2nd order polynominal for the lineshape of the NR$(\jpsi\antiproton)$ contribution. Moreover, a model with $J^P=1/2^+$ assignment to the ${P_{\psi s}^{\Lz}(4338)}^0$ state is also considered.
Finally, the behaviour of the maximum-likelihood estimator is studied using 1000 pseudoexperiments. Biases on the fit parameters are present due to the limited sample size and are assigned as systematic uncertainties.  For the reconstruction efficiency, 
the nominal efficiency function based on decays from either the long  or downstream \Lz category and the largest effect is considered as systematic uncertainty. 

Additional systematic uncertainties account for the limited knowledge of the  $\Lz \to\proton\pim$ decay amplitude parameters~\cite{BESIII:2018cnd,PDG2022}, the background parameterization, and the effect of the resolution on the $m(\jpsi\Lz)$ invariant mass. 
The nominal background parameterization ${\cal P}_{\rm{bkg}}$ is obtained from the distributions of candidates in the $m(\jpsi\Lz\antiproton)$ range ${[5200,5250] \cup [5340,5350]\mev}$, while the parameterization obtained from the region $[5295, 5315]\mev$ is used to assess systematic effects. The background fraction $\beta=0.07\pm0.01$ is also varied within uncertainties. The effect of the invariant mass resolution, about $1 \mev$ on average on $m(\jpsi\Lz)$, is estimated by smearing the invariant mass distributions of 1000 pseudoexperiments and fitting them using the nominal model.
\begin{table}[tb]
\centering
\caption{\label{tab:syst} Systematic uncertainties on the  mass ($M_{P_{\psi s}^{\Lz}}$) and width ($\Gamma_{P_{\psi s}^{\Lz}}$) of the ${P_{\psi s}^{\Lz}}^0$ state (in \mev), and on the fit fractions  $f_{P_{\psi s}^{\Lz}}$,  $f_{\textrm{NR} (\jpsi\antiproton)}$ and $f_{\textrm{NR} (\Lz\antiproton)}$ of the pentaquark candidate and nonresonant contributions (in \%).}
\begin{tabular}{lccccc}
\hline
  Source         & $M_{P_{\psi s}^{\Lz}}$    & $\Gamma_{P_{\psi s}^{\Lz}}$   & $f_{P_{\psi s}^{\Lz}}$  & $f_{\textrm{NR} (\jpsi\antiproton)}$  & $f_{\textrm{NR} (\Lz\antiproton)}$\\
  \hline Hadron radius                 & $0.1$       & $0.4$   &  $0.3$  & $0.2$  & $0.2$ \\
 $LS$ values           & $0.3$       & $0.1$   &  $0.8$  & $0.7$  & $0.6$\\
 Breit--Wigner ${\overline{P}_{\psi}^{N}}^-$      & $0.1$       & $0.9$   &  $0.8$  & ... & ...\\
  $J^P({P_{\psi s}^{\Lz}}^0)$ assignment    & $0.1$       & $0.9$   &  $1.2$  & $0.4$  & $0.9$  \\
  Fitting procedure                    & $0.1$       & $0.2$   &  $0.1$  & $1.0$  & 
  $1.1$ \\
  Efficiency           & $0.02$       & $0.19$   &  $0.02$  & $0.3$  & $0.2$  \\
$\Lz$ decay parameters  & $0.02$       & $0.04$   &  $0.01$  & $0.3$  & $0.2$  \\
   Background &    $0.01$     & $0.05$   &  $0.96$  & $0.4$  & $0.7$  \\
  Mass resolution             & $0.01$       & $0.03$   &  $0.01$  & $0.1$  & $0.1$  \\
  \hline
  Total                       & $0.4$       & $1.3$   &  $1.9$  & $1.4$  &  $1.7$ \\
\hline
\end{tabular}
\end{table}

The mass and width of the new pentaquark candidate are measured to be ${M_{P_{\psi s}^{\Lz}}= 4338.2\pm 0.7 \pm 0.4 \mev}$ and $\Gamma_{P_{\psi s}^{\Lz}}=7.0\pm1.2\pm1.3\mev$; the measured fit fractions are $f_{P_{\psi s}^{\Lz}}=0.125 \pm 0.007 \pm 0.019$, $f_{{\rm NR} (\jpsi\antiproton)}=0.840 \pm 0.022 \pm 0.014$, and  $f_{{\rm NR} (\Lz\antiproton)}=0.113 \pm 0.013 \pm 0.017$ for the resonant ${P_{\psi s}^{\Lz}}^0$ state, the nonresonant ${\textrm{NR} (\jpsi\antiproton)}$, and  ${\textrm{NR} (\Lz\antiproton)}$ contributions, respectively. The first uncertainty is statistical and the second systematic. The $J^P=1/2^-$ quantum numbers for the ${P_{\psi s}^{\Lz}(4338)}^0$ state are preferred; $J=1/2$ is established and positive parity can be excluded at 90\% confidence level. 

Because of the small $Q$-value of the decay, the most precise single measurement to date of the \Bm mass $5279.44 \pm 0.05 \pm 0.07\mev$ is performed. This measurement is based on 1670 signal candidates with \Lz baryons in the long category, which amounts to 36\% of the total. Systematic uncertainties on the \Bm mass include uncertainties on particle interactions
with the detector material ($0.030 \mev$), momentum scaling due to imperfections in the magnetic-field mapping ($0.039\mev$)~\cite{LHCb-DP-2014-002}, and the choice of the signal and background fit model ($0.050\mev$). The alternative fit model, with compatible fit quality with respect to the nominal model, comprises an exponential function for the background and a sum of a Gaussian and two Johnson functions for the signal.  Systematic uncertainties from knowledge of the $\jpsi, \Lz$, and $\proton$ masses are negligible. 

In conclusion, an amplitude analysis of the $\Bm\to\jpsi\Lz\antiproton$ decay is performed using about 4400 signal candidates selected on data collected by the \lhcb experiment between 2011 and 2018 and corresponding to an integrated luminosity of $9 \invfb$. A new resonant structure in the $\jpsi\Lz$ system is found with high statistical significance, representing the first observation of a pentaquark candidate with strange quark content named the ${P_{\psi s}^{\Lz}(4338)}^0$ state, with spin $J=1/2$ assigned and parity $P=-1$ preferred. The new ${P_{\psi s}^{\Lz}(4338)}^0$ state is found at the threshold for $\Xicp\Dm$ baryon-meson production, which is relevant for the interpretation of its nature. No evidence for additional resonant states, either ${P_{\psi s}^{\Lz}(4255)}^0$ or ${\overline{P}_{\psi}^{N}}^-$ pentaquark candidates or excited $K^{-}$ resonances, is found from the fit to data.

%% file: acknowledgements.tex
\section*{Acknowledgements}
%
%
\noindent We express our gratitude to our colleagues in the CERN
accelerator departments for the excellent performance of the LHC. We
thank the technical and administrative staff at the LHCb
institutes.
We acknowledge support from CERN and from the national agencies:
CAPES, CNPq, FAPERJ and FINEP (Brazil); 
MOST and NSFC (China); 
CNRS/IN2P3 (France); 
BMBF, DFG and MPG (Germany); 
INFN (Italy); 
NWO (Netherlands); 
MNiSW and NCN (Poland); 
MEN/IFA (Romania); 
MICINN (Spain); 
SNSF and SER (Switzerland); 
NASU (Ukraine); 
STFC (United Kingdom); 
DOE NP and NSF (USA).
We acknowledge the computing resources that are provided by CERN, IN2P3
(France), KIT and DESY (Germany), INFN (Italy), SURF (Netherlands),
PIC (Spain), GridPP (United Kingdom), 
CSCS (Switzerland), IFIN-HH (Romania), CBPF (Brazil),
Polish WLCG  (Poland) and NERSC (USA).
We are indebted to the communities behind the multiple open-source
software packages on which we depend.
Individual groups or members have received support from
ARC and ARDC (Australia);
Minciencias (Colombia);
AvH Foundation (Germany);
EPLANET, Marie Sk\l{}odowska-Curie Actions and ERC (European Union);
A*MIDEX, ANR, IPhU and Labex P2IO, and R\'{e}gion Auvergne-Rh\^{o}ne-Alpes (France);
Key Research Program of Frontier Sciences of CAS, CAS PIFI, CAS CCEPP, 
Fundamental Research Funds for the Central Universities, 
and Sci. \& Tech. Program of Guangzhou (China);
GVA, XuntaGal, GENCAT and Prog.~Atracci\'on Talento, CM (Spain);
SRC (Sweden);
the Leverhulme Trust, the Royal Society
 and UKRI (United Kingdom).

%% file: appendix.tex

\section*{Supplemental Material}

\appendix

\section{Amplitude model}
\label{app:Amplitude_model}

The amplitude model is constructed using  helicity formalism~\cite{Chung:186421} following the prescription for final particle spin matching described in Ref.~\cite{DPdeco}.
The amplitude $O^{X}_{\lambda_1,\lambda_2,\lambda_3}$ describes the decay amplitude for the \Bm to the \jpsi \Lz \antiproton final state via the $K^{*-},{\overline{P}_{\psi}^{N}}^-,{P_{\psi s}^{\Lz}}^0$ decay chains as follows,
\begin{align} 
  O^{K^{*}}_{\lambda_\jpsi,\lambda_{\Lz},\lambda_\antiproton}(m^2_{\jpsi {\Lz}},m^2_{{\Lz} \antiproton})
  \nonumber
  &= \sum_{j^{K^{*}}} \sum_{\{\lambda'\}}
            \sqrt{\frac{2j^{K^{*}}+1}{4\pi}}\,
             \HellH{B^-\to K^{*}\jpsi}_{\lambda_\jpsi'}\,
            R(m^2_{{\Lz} \antiproton})\,
            d^{j^{K^{*}}}_{\lambda_\jpsi',\lambda_{\Lz}'-\lambda_\antiproton'}(\theta_{K^{*}}) \\ \nonumber
            & \hspace{-1.2cm} \times
           \HellH{K^{*}\to {\Lz} \antiproton}_{\lambda_{\Lz}',\lambda_\antiproton'} \delta_{\lambda_\jpsi',\lambda_\jpsi}\,
            d^{1/2}_{\lambda_{\Lz}',\lambda_{\Lz}}(\wigner^{{\Lz}}_{B\antiproton})\,
            d^{1/2}_{\lambda_\antiproton',\lambda_\antiproton}(-\wigner^{\antiproton}_{B{\Lz}}) \times (-1)^{j^{\jpsi}-\lambda_\jpsi'}(-1)^{j^\antiproton-\lambda_\antiproton'},\\
  O^{P_{\psi}^{N}}_{\lambda_\jpsi,\lambda_{\Lz},\lambda_\antiproton}(m^2_{\jpsi {\Lz}},m^2_{{\Lz}\antiproton})
  \nonumber
  &= \sum_{j^{P_{\psi}^{N}}} \sum_{\{\lambda'\}}
            \sqrt{\frac{2j^{P_{\psi}^{N}}+1}{4\pi}}\,
           \HellH{B^-\to P_{\psi}^{N}{\Lz}}_{\lambda_{\Lz}'}\,
            R(m^2_{\antiproton \jpsi})\,
            d^{j^{P_{\psi}^{N}}}_{\lambda_{\Lz}',\lambda_\antiproton'-\lambda_\jpsi'}(\theta_{P_{\psi}^{N}})\\  
            & \nonumber \hspace{-2.cm}\times \HellH{P_{\psi}^{N}\to \antiproton\jpsi}_{\lambda_\antiproton',\lambda_\jpsi'}  d^{1}_{\lambda_\jpsi',\lambda_\jpsi}(-\wigner^{\jpsi}_{B\antiproton})\,
            \delta_{\lambda_{\Lz}',\lambda_{\Lz}}\,
            d^{1/2}_{\lambda_\antiproton',\lambda_\antiproton}(\wigner^{\antiproton}_{B\jpsi}) \times (-1)^{j^{\Lz}-\lambda_\Lz'}(-1)^{j^\jpsi-\lambda_\jpsi'},\\
  O^{{{P}_{\psi s}^{\Lz}}}_{\lambda_\jpsi,\lambda_{\Lz},\lambda_\antiproton}(m^2_{\jpsi {\Lz}},m^2_{{\Lz} \antiproton})
  \nonumber
  &= \sum_{j^{{P}_{\psi s}^{\Lz}}} \sum_{\{\lambda'\}}
     \sqrt{\frac{2j^{{P}_{\psi s}^{\Lz}}+1}{4\pi}}\,
            \HellH{B^-\to {P}_{\psi s}^{\Lz} \antiproton}_{\lambda_\antiproton'}\,
            R(m^2_{\jpsi{\Lz}})\,
            d^{J^{{P}_{\psi s}^{\Lz}}}_{\lambda_\antiproton,\lambda_\jpsi'-\lambda_{\Lz}'}(\theta_{{P}_{\psi s}^{\Lz}}) \\ \nonumber
            & \hspace{-1.2cm} \times
            \HellH{{P}_{\psi s}^{\Lz}\to \jpsi{\Lz}}_{\lambda_\jpsi',\lambda_{\Lz}'} d^{1}_{\lambda_\jpsi',\lambda_\jpsi}(\wigner^{\jpsi}_{B{\Lz}})\,
            d^{1/2}_{\lambda_{\Lz}',\lambda_{\Lz}}(-\wigner^{{\Lz}}_{B\jpsi})\,
            \delta_{\lambda_\antiproton',\lambda_\antiproton} \times (-1)^{j^{\antiproton}-\lambda_\antiproton'}(-1)^{j^{\Lz}-\lambda_\Lz'},
            \nonumber\\
\end{align} 
where $j^X$ is the total angular momentum of the different contributions in the ${X=K^{*-}, \overline{P}_{\psi}^{N-}}$ and ${P_{\psi s}^{\Lz}}^0$ decay chains, respectively, and $\{\lambda'\}$ are the helicities of the final particles before spin rotations. The angle, $\wigner^i_{Bk}$, is between the $B^-$ and the particle $k$ in rest frame $i$. The coupling, $H_{\lambda'}^{A\to BC}$, is the helicity coupling of a two-body decay $A\to BC$, $R$ is the line shape and $d^j_{\lambda_A,\lambda_B-\lambda_C}$ is the small Wigner function.
The angle, $\theta_X$, is the helicity angle of particle $X$, 
which is calculated using the \Lz in the $K^{*-}$ rest frame, and either the \antiproton in the $\overline{P}_{\psi}^{N-}$ rest frame, or the \jpsi in the $P_{\psi s}^{\Lz}$ rest frame. 

The total decay amplitude is obtained by including the $\jpsi\to \mup\mu^-$ and the $\Lz\to\proton\pim$ decay amplitudes
\begin{align}
    A_{\lambda_\antiproton,\lambda_{p},\Delta\mu}(m_{\antiproton\Lz},\vec{\Omega}) &= \sum_{\lambda_{\bar \Lambda},\lambda_\jpsi}\left(O^{K^{*-}} + O^{P_{\psi}^{N-}} + O^{{P_{\psi s}^{\Lz}}^0}\right)_{\lambda_\jpsi,\lambda_{\Lz},\lambda_\antiproton}(m^2_{\jpsi {\Lz}},m^2_{{\Lz} \antiproton})\nonumber\\
    & \hspace{2cm}\times D^{1*}_{\lambda_\jpsi,\Delta_\mu}(\phi_{\mu^-},\theta_\jpsi,0)\,
    \HellH{{\Lz}\to {p}\pi}_{\lambda_{p}}D^{1/2*}_{\lambda_{\Lz},\lambda_{ p}}(\phi_{p},\theta_{\Lz},0),
    \label{eq:tot_amp}
\end{align}
where $D^{j*}_{\lambda_A, \lambda_B-\lambda_C}(\phi, \theta, 0)$ is the Wigner $D$ matrix, equal to $e^{i\lambda_A\phi}d^j_{\lambda_A, \lambda_B-\lambda_C}(\theta)$,  $\phi_{\mu^-},\theta_\jpsi,\phi_{p},\theta_{\Lz}$ are the azimuthal and polar angles of $\mu^-$ and \proton in the $\jpsi$ and $\Lz$ rest frames, respectively. The axes in the \B rest frame are defined as follows,
\begin{align}
    &\begin{aligned}\hat x_\jpsi &= \hat y_\jpsi \times \hat z_\jpsi,\\
    \hat y_\jpsi &= \frac{\vec p_\antiproton^B \times \vec p_\jpsi^B}{|\vec p_\jpsi^B \times \vec p_\antiproton^B|},\\ 
    \hat z_\jpsi &= \frac{\vec p_\jpsi^B}{|\vec p_\jpsi^B|}, \\ 
    \end{aligned}
    &\begin{aligned}\hat x_{\Lz} &= \hat y_{\Lz} \times \hat z_{\Lz},\\
    \hat y_{\Lz} &= \frac{\vec p_\antiproton^B \times \vec p_{ \Lz}^B}{|\vec p_{\Lz}^B \times \vec p_\antiproton^B|},\\ 
    \hat z_{\Lz} &= \frac{\vec p_{\Lz}^B}{|\vec p_{\Lz}^B|}, 
    \end{aligned}
\end{align}
where the symbol $\hat x$ refers to $\vec x / |x|$. 
In Eq.~\ref{eq:tot_amp}, $\Delta\mu$ is the difference of the muon helicities. For the $\jpsi\to \mup\mu^-$ decay, the coupling can be absorbed into the other couplings of the total decay amplitude and therefore is not fit. Indeed, there is only one coupling because the process with $\Delta\mu=0$ is highly suppressed. So, $\Delta\mu$ can only take values $1$ and $-1$, and both choices lead to the same helicity coupling due to parity conservation.

Enforcing \CP conservation, the helicity couplings for $B^-$ and \Bp decays are the same.
The matrix-element formula is the same for charge-conjugate decays, but all azimuthal angles must change sign due to charge-parity transformation, \ie  $\phi_{p} \to -\phi_{\antiproton}$ and $\phi_{\mu^-}\to -\phi_{\mu^+}$.

The $\Lbar \to \antiproton \pip$ decay parameters are defined by
\begin{align}
\label{eq:Lpar}
      \alpha_{+} = \frac{|H^{\Lbar \to \antiproton \pip}_{1/2}|^2 - |H^{\Lbar \to \antiproton \pip}_{-1/2}|^2}{|H^{\Lbar \to \antiproton \pip}_{1/2}|^2 + |H^{\Lbar \to \antiproton \pip}_{-1/2}|^2}, \nonumber \\
      \beta_{+} = \frac{2\textrm{Im}(H^{\Lbar \to \antiproton \pip}_{1/2}H^{\Lbar \to \antiproton \pip *}_{-1/2})}{|H^{\Lbar \to \antiproton \pip}_{1/2}|^2 + |H^{\Lbar \to \antiproton \pip}_{-1/2}|^2}, \\
      \gamma_{+} = \frac{2\textrm{Re}(H^{\Lbar \to \antiproton \pip}_{1/2}H^{\Lbar \to \antiproton \pip *}_{-1/2})}{|H^{\Lbar \to \antiproton \pip}_{1/2}|^2 + |H^{\Lbar \to \antiproton \pip}_{-1/2}|^2}, \nonumber
\end{align}
which satisfy the relation,
\begin{equation}
    \alpha^2_{+} + \beta^2_{+} + \gamma^2_{+} =1.
    \nonumber
\end{equation}

It is convenient to express $\beta_{+}$ and $\gamma_{+}$ in terms of an angle $\phi_{+}$ defined as
\begin{equation}
  \begin{aligned}
      \beta_{+}  = \sqrt{1-\alpha^2_{+}}\sin{\phi_{+}}, \\
      \gamma_{+} = \sqrt{1-\alpha^2_{+}}\cos{\phi_{+}}. \\
  \end{aligned}
\end{equation}

Enforcing \CP conservation, the following relations hold,
\begin{align}
H^{\Lbar\to\antiproton\pip}_{\mp 1/2}=\eta_{\Lz}\eta_{\proton}\eta_{\pi}(-1)^{J_{\Lz}-J_{\proton}-J_{\pi}}H^{\Lz\to\proton\pim}_{\pm 1/2}=-H^{\Lz\to\proton\pim}_{\pm 1/2}.
\end{align} 

This leads to
\begin{align}
      \alpha_{+} = -\alpha_{-}, \nonumber\\
      \beta_{+}  = -\beta_{-}, \nonumber\\
      \gamma_{+} = \gamma_{-}, \nonumber\\
      \tan{\phi_{+}} = -\tan{\phi_{-}},
\end{align}
where $\alpha_{-}$, $\beta_{-}$ and $\gamma_{-}$ are obtained following Eq.~\ref{eq:Lpar} but using the couplings of the conjugate decay. 

If the complex helicity coupling $H^{\LbarTopbarpip}_{1/2}$ is set to (1,0), then $H^{\LbarTopbarpip}_{-1/2}=\sqrt{\frac{1-\alpha_{+}}{1+\alpha_{+}}}e^{-i\phi_{+}}$.
The values of $\alpha_{+}$ and $\phi_{+}$ are fixed to $-0.758$ \cite{BESIII:2018cnd} and $+6.5^{o}$ \cite{PDG2022} respectively.

The helicity couplings for the decay $A\to B C$ can be expressed as a combination of the $LS$ couplings ($B_{L,S}$) using the Clebsch--Gordan (CG) coefficients
\begin{align}
    H^{A\to BC}_{\lambda_B,\lambda_C} = \sum_{L}\sum_{S}\sqrt{\frac{2L+1}{2J_{A}+1}}
    B_{L,S}&\braket{{J_{B}}, {\lambda_{B}}, {J_{C}},   {-\lambda_{C}}}{ {S}, {\lambda_{B}-\lambda_{C}}} \nonumber\\ \times &\braket{{L}, {0}, {S},{\lambda_{B}-\lambda_{C}}}{ {J_A}, {\lambda_B-\lambda_C}},
\end{align}
where $L$ is the orbital angular momentum in the decay, and $S$ is the total spin of the daughters, $\vec{S}=
\vec{J}_B + \vec{J}_C$ ($|J_B-J_C|\leq S \leq J_B + J_C$).
If the $Q$ value, defined as $Q=M_A - M_B - M_C$, is small ($Q/M_A \ll 1$), the higher orbital angular momenta are suppressed, hence the number of couplings is reduced.
CG coefficients automatically take into account parity conservation constraints on helicity couplings for a strong or electromagnetic decay.

For $\Bm\to XR$, $R\to YZ$ cascade decay, \eg $X=\antiproton$, $R={P_{\psi s}^{\Lz}}^0$,  $Y=\Lz$ and $Z=\jpsi$, the lineshape of $R$ is
\begin{equation}
  \begin{aligned}
    \left(\frac{p}{p_0}\right)^L B^\prime_L(p,p_0,d) \times \left(\frac{q}{q_0}\right)^l B^\prime_l(q,q_0,d) \textrm{BW}(m|m_0,\Gamma_0),
  \end{aligned}
\end{equation}
where $p$ is the momentum of resonance $R$ in the \Bm rest frame, 
$q$ is the  momentum of particle $Y$ in the rest frame of resonance $R$,
$p_0$ and $q_0$ are the momentum values calculated at the $R$ resonance peak,
$L$ is the orbital angular momentum between resonance $R$ and particle $X$ in the $\Bm\to XR$ decay, and
$l$ is the orbital angular momentum between particle $Y$ and particle $Z$ in the $R\to YZ$ decay.
The $\left(p/p_0\right)^L$ and $\left(q/q_0\right)^l$ contributions are the orbital barrier factors,
$B^\prime_L(p,p_0,d)$ and $B^\prime_l(q,q_0,d)$ are the Blatt--Weisskopf functions that account for the difficulty to create the orbital angular momentum, and depend on the production (decay) momentum $p$ ($q$) and on the size of the decaying particle given by the hadron radius $d$. These coefficients up to order 4 are listed below,
\begin{align}
&B_{0}^{\prime}\left(p, p_{0}, d\right)=1,\nonumber\\ \nonumber
&B_{1}^{\prime}\left(p, p_{0}, d\right)=\sqrt{\frac{1+\left(p_{0} d\right)^{2}}{1+(p d)^{2}}},\\ \nonumber
&B_{2}^{\prime}\left(p, p_{0}, d\right)=\sqrt{\frac{9+3\left(p_{0} d\right)^{2}+\left(p_{0} d\right)^{4}}{9+3(p d)^{2}+(p d)^{4}}},\\ \nonumber 
&B_{3}^{\prime}\left(p, p_{0}, d\right)=\sqrt{\frac{225+45\left(p_{0} d\right)^{2}+6\left(p_{0} d\right)^{4}+\left(p_{0} d\right)^{6}}{225+45(p d)^{2}+6(p d)^{4}+(p d)^{6}}},\\ 
&B_{4}^{\prime}\left(p, p_{0}, d\right)=\sqrt{\frac{11025+1575\left(p_{0} d\right)^{2}+135\left(p_{0} d\right)^{4}+10\left(p_{0} d\right)^{6}+\left(p_{0} d\right)^{8}}{11025+1575(p d)^{2}+135(p d)^{4}+10(p d)^{6}+(p d)^{8}}},
\label{eq:BlattWeisk}
\end{align}
where $d$ is the particle size parameter, set to $3 \gev^{-1}$ following the convention of Ref.~\cite{LHCb-PAPER-2015-029}. 
In the nominal amplitude fit of $\Bm\to\jpsi\Lz\antiproton$ decays, the constant $d$ is set to $d_{B}=d_{R}=3$ \gev$^{-1}$ for the \Bm and intermediate resonant $R$ decays.

The relativistic Breit--Wigner amplitude is given by
\begin{equation}
  \begin{aligned}
    \textrm{BW}(m|m_0,\Gamma_0) = \frac{1}{m^2_0 - m^2 - im_0\Gamma(m)},
  \end{aligned}
\end{equation}
with 
\begin{equation}
  \begin{aligned}
    \Gamma(m)=\Gamma_0\left(\frac{q}{q_0}\right)^{2l+1}\left(\frac{m_0}{m}\right)B^\prime_l\left(q,q_0,d\right)^2,\label{eq:width}
  \end{aligned}
\end{equation}
where $m$ is the invariant mass of the $YZ$ system, and $m_0$ ($\Gamma_0$) is the mass (width) of the $R$ resonance.
In the case that resonance $R$ has a mass peak outside of the accessible kinematic region, \ie $m_R>m_\Bm-m_X$, such as for the $K_2(2250)^{-}$ and $K_3(2320)^{-}$ states, the effective mass $m^{\textrm{eff}}_0$ 
is introduced to calculate the two-body-decay momentum $q_0$ in Eq.~\ref{eq:width},
\begin{equation}
  \begin{aligned}
    m^{\textrm{eff}}_0(m_0)=m^{\textrm{min}} +\frac{1}{2}(m^{\textrm{max}}-m^{\textrm{min}})\left[ 1 + \tanh\left(\frac{m_0-\frac{m^{\textrm{min}}+m^{\textrm{max}}}{2}}{m^{\textrm{max}}-m^{\textrm{min}}}\right)\right].
  \end{aligned}
\end{equation}
This term is a constant that can be absorbed into the couplings, since it enters only in Eq.~\ref{eq:width}, and the mass $m_0$ and width $\Gamma_0$ of the $K^*$ resonant contributions are fixed to the nominal values~\cite{PDG2022}. 
In the case of a resonance $R$ with mass peak located outside of the phase space at values $m_R<m_Y+m_Z$, such as for the $K_4(2045)^{-}$ state, the width is chosen as mass-independent parameter $\Gamma_0$.  
In the nominal model, the non-resonant (NR) contribution is modelled by a second-order polynomial,
\begin{equation}
      c_0 + c_1(m-m_0) + c_2(m-m_0)^2,
\end{equation}
where $m_0$ is the average value of the invariant mass distribution, \ie of the $m_{\jpsi\antiproton}$ invariant mass distribution. The coefficients, $c_i$, are the polynomial coefficients, where $c_0$ is set to a constant value since one of the $c_{i}$ coefficients can be factor out of amplitude matrix element, and the other two are extracted from a fit to the data.  
\section{Event-by-event efficiency parameterisation}
Event-by-event acceptance corrections are applied to the data using an efficiency parameterisation based on the decay kinematics. The 6-body phase space of the topology $\Bm\to\jpsi(\to\mun \mup)\Lz(\to\proton\pim)\antiproton$ is fully described by six independent kinematic variables: $m_{\Lz\antiproton}$, $\cos\theta_{K^*}$, $\cos\theta_{J/\psi}$, $\phi_{\mu}$, $\cos\theta_{\Lbar}$, and $\phi_{\antiproton}$. For the signal mode, the overall efficiency, including trigger, detector acceptance, and selection procedure, is obtained from simulation as a function of the six kinematic variables, $\vec{\omega} \equiv \{\cos\theta_{K^*}, \cos\theta_{\jpsi}, \phi'_{\mu}, m'_{\Lz\antiproton}, \cos\theta_{\Lbar}, \phi'_{\antiproton}\}$. Here, $m'_{\Lz\antiproton}$ and $\phi'$ are transformed such that all four variables in $\vec{\omega}$ lie in the range $(-1,1]$. The efficiency is parameterised as the product of Legendre polynomials
\begin{equation}
  \begin{aligned}
    \epsilon(\vec{\omega}) = \sum_{i,j,k,l,m,n} & c_{i,j,k,l,m,n}
    P(\cos\theta_{K^*},i)P(\cos\theta_{J/\psi},j) \\
   &  P(\phi'_{\mu},k)P(m'_{\Lz\antiproton},l)
    P(\cos\theta_{\Lbar},m)P(\phi'_{\antiproton},n),
  \end{aligned}
\end{equation}
where $P(x,l)$ are Legendre polynomials of order $l$ in $x\in(-1,1]$. Employing the order of the polynomials as $\{2,2,2,2,4,3\}$ for $\{\cos\theta_{K^*}, \cos\theta_{\jpsi}, \phi'_{\mu}, m'_{\Lz\antiproton}, \cos\theta_{\Lbar}, \phi'_{\antiproton}\}$, respectively, was found to give a good parameterisation.
The coefficients $c_{i,j,k,l,m,n}$ are determined from a moment analysis of $\Bm\to\jpsi\Lz\antiproton$ phase-space simulated samples
\begin{equation}
  \begin{aligned}
    c_{i,j,k,l,m,n}=\frac{C}{\sum_{\nu}\omega_{\nu}}\sum^{N_{rec}}_{\nu=1}\omega_{\nu}\left(\frac{2i+1}{2}\right)\left(\frac{2j+1}{2}\right)\left(\frac{2k+1}{2}\right)\left(\frac{2l+1}{2}\right)\left(\frac{2m+1}{2}\right)\left(\frac{2n+1}{2}\right) \\
    \times P(\cos\theta_{K^*},i)P(\cos\theta_{J/\psi},j)P(\phi'_{\mu},k)P(m'_{\Lz\antiproton},l)P(\cos\theta_{\Lbar},m)P(\phi'_{\antiproton},n),
  \end{aligned}\label{eqn:eff_coeff_mom}
\end{equation}
where $\omega_{\nu}$ is the per-event weight taking into account both the generator-level phase-space element, $\deriv\Phi$, and the kinematic event weights. Simulation samples are employed where $\Bm\to\jpsi\Lz\antiproton$ events are generated uniformly in phase space. In order to render the simulation flat also in $m(\Lz\antiproton)$, the inverted phase-space factor, $1/\deriv\Phi$, is considered. The factors of $(2a + 1)/2$ arise from the orthogonality of the Legendre polynomials,
\begin{equation}
\int_{-1}^{+1} P(x, a) P(x, a') \deriv x = \frac{2}{2 a +1}\delta_{ a a'}  ~.
\end{equation}
The sum in Eq. \ref{eqn:eff_coeff_mom} is over the reconstructed events in the simulation sample after all selection criteria. 
The factor $C$ ensures appropriate normalisation and it is computed such that
\begin{align} 
\sum_{n=0}^{N_{\rm gen}} \varepsilon(\vec{x}_n) = N_{\rm rec},
\end{align} 
where $N_{\rm rec}$ is the total number of reconstructed signal events. 

Up to statistical fluctuations, the parameterisation follows the simulated data in all the distributions.

\section{Fit results of the nominal model}
In Table~\ref{amfit:tab:paraPcsNRpJpsi}, the fit results of the nominal model are reported including the results of the $LS$ couplings. 
The couplings are split into real and imaginary parts, \ie Re$_\textrm{prod(decay)}(R)_{L,S}$, Im$_\textrm{prod(decay)}(R)_{L,S}$. The subscript $\textrm{prod}$ ($\textrm{decay}$) refers to the $\Bm\to X R$ ($R\to YZ$) process, where $X$, $Y$, $Z$ are the final state particles, and $R$ is the decay chain under consideration. The subscript $L$ refers to the orbital angular momentum and $S$ to the sum of the spins of the decay products. 

\begin{table}[ht]
\centering
\small
\caption{\small\label{amfit:tab:paraPcsNRpJpsi}Parameters determined from the fit to data using the nominal model where uncertainties are statistical only.}
\begin{tabular}{lc}
\hline
Parameters & Values \\
\hline
$M_{P_{\psi s}^{\Lz}}$ (\mev) & $ 4338.2 \pm 0.7  $ \\
$\Gamma_{P_{\psi s}^{\Lz}}$ (\mev) & $7.0\pm1.0$ \\
Re$_{\textrm{decay}}$(${P}_{\psi s}^{\Lz \ 0}$)$_{L=0,S=1/2}$ & $0.16\pm0.04$ \\
Im$_{\textrm{decay}}$(${P}_{\psi s}^{\Lz \ 0}$)$_{L=0,S=1/2}$ & $-0.04\pm0.08$ \\
  Re$_{\textrm{decay}}$(NR(\antiproton\jpsi))$_{L=1,S=3/2}$ & $-3.0\pm0.4$ \\
  Im$_{\textrm{decay}}$(NR(\antiproton\jpsi))$_{L=1,S=3/2}$ & $0.1\pm0.5$ \\
  Re$_{\textrm{prod}}$(NR(\antiproton\Lz))$_{L=1,S=1}$ & $1.0\pm0.5$ \\
Im$_{\textrm{prod}}$(NR(\antiproton\Lz))$_{L=1,S=1}$ & $-0.5\pm0.5$ \\
Re$_{\textrm{prod}}$(NR(\antiproton\Lz))$_{L=2,S=2}$ & $0.01\pm0.25$ \\
Im$_{\textrm{prod}}$(NR(\antiproton\Lz))$_{L=2,S=2}$ & $0.1\pm0.4$ \\
Re$_{\textrm{decay}}$(NR(\antiproton\Lz))$_{L=0,S=1}$ & $-0.3\pm0.1$ \\
Im$_{\textrm{decay}}$(NR(\antiproton\Lz))$_{L=0,S=1}$ & $-0.1\pm0.1$ \\
Re$_{\textrm{decay}}$(NR(\antiproton\Lz))$_{L=2,S=1}$ & $0.4\pm0.1$ \\
Im$_{\textrm{decay}}$(NR(\antiproton\Lz))$_{L=2,S=1}$ & $0.1\pm0.2$ \\
$c_{1}$ & $2.6\pm0.6$ \\
$c_{2}$ & $72\pm14$ \\
\hline
  $f_\textrm{$P_{\psi s}^{\Lz}$}$ & $0.125\pm0.007$ \\
  $f_{\textrm{NR}(\antiproton\jpsi)}$ & $0.840\pm0.002$ \\
  $f_{\textrm{NR}(\antiproton\Lz)}$ & $0.113\pm0.013$ \\
 \hline
$-\log{L}$ & $-807.63$  \\
\hline  
\end{tabular}
\end{table}

\section{Angular moments}
The normalized angular moments $\left<P_j^U\right>$ of the ${{P}_{\psi s}^{\Lz}}^0$ helicity angle are defined as,
\begin{equation}
    \left<P_j^U\right> = \sum_{i=0}^{N_{rec}} \omega_i P_j\left(\cos\theta_{{P}_{\psi s}^{\Lz}}\right)
\end{equation}
where $N_{rec}$ is the number of  selected events, $P_j$ are Legendre polynomials and $\omega_i$ are per-event weights accounting for background subtraction (with \sPlot technique) and efficiency correction.  

\begin{figure}[tb]
\centering
\small
\includegraphics[width=0.32\textwidth]{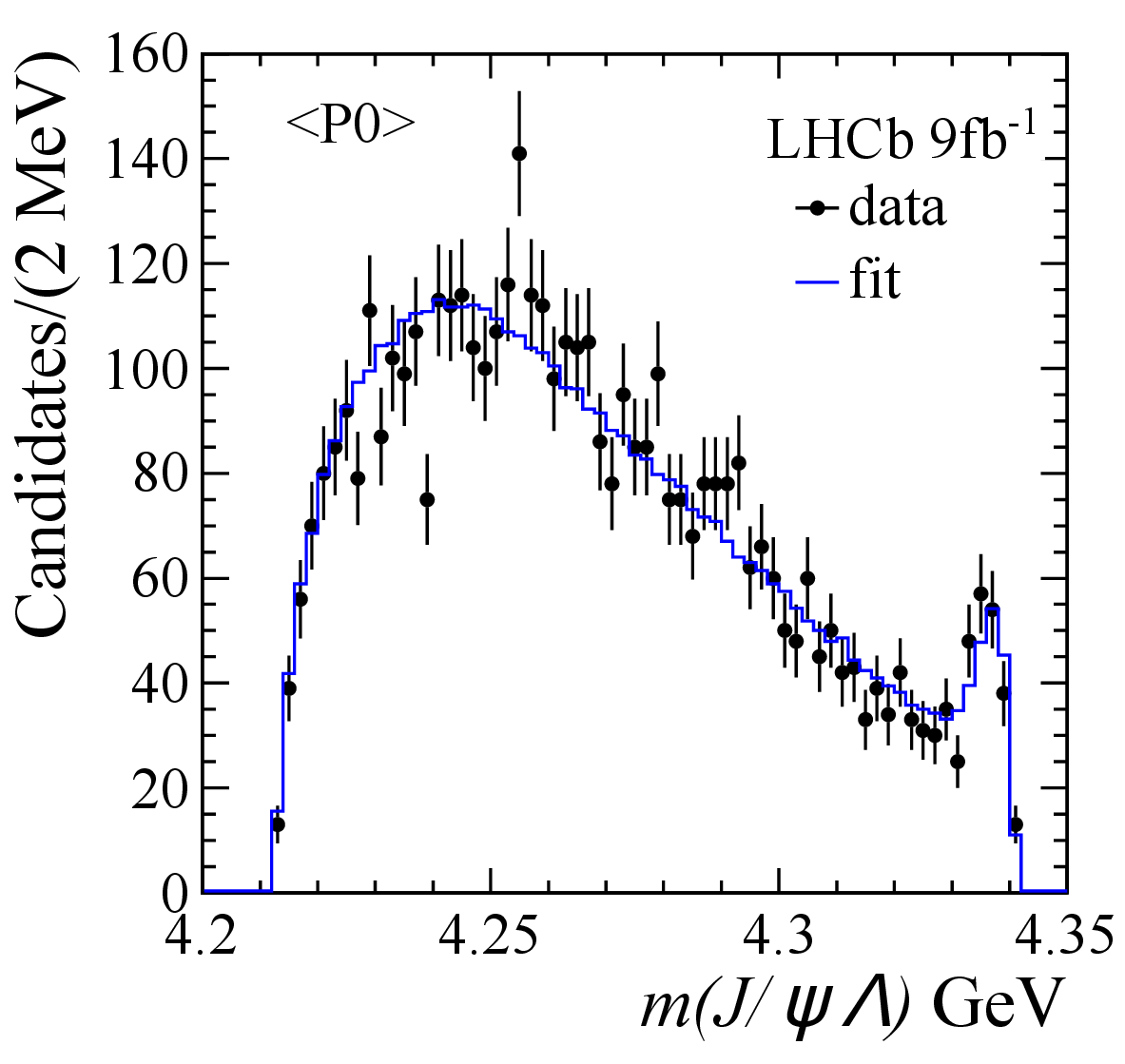}
\includegraphics[width=0.32\textwidth]{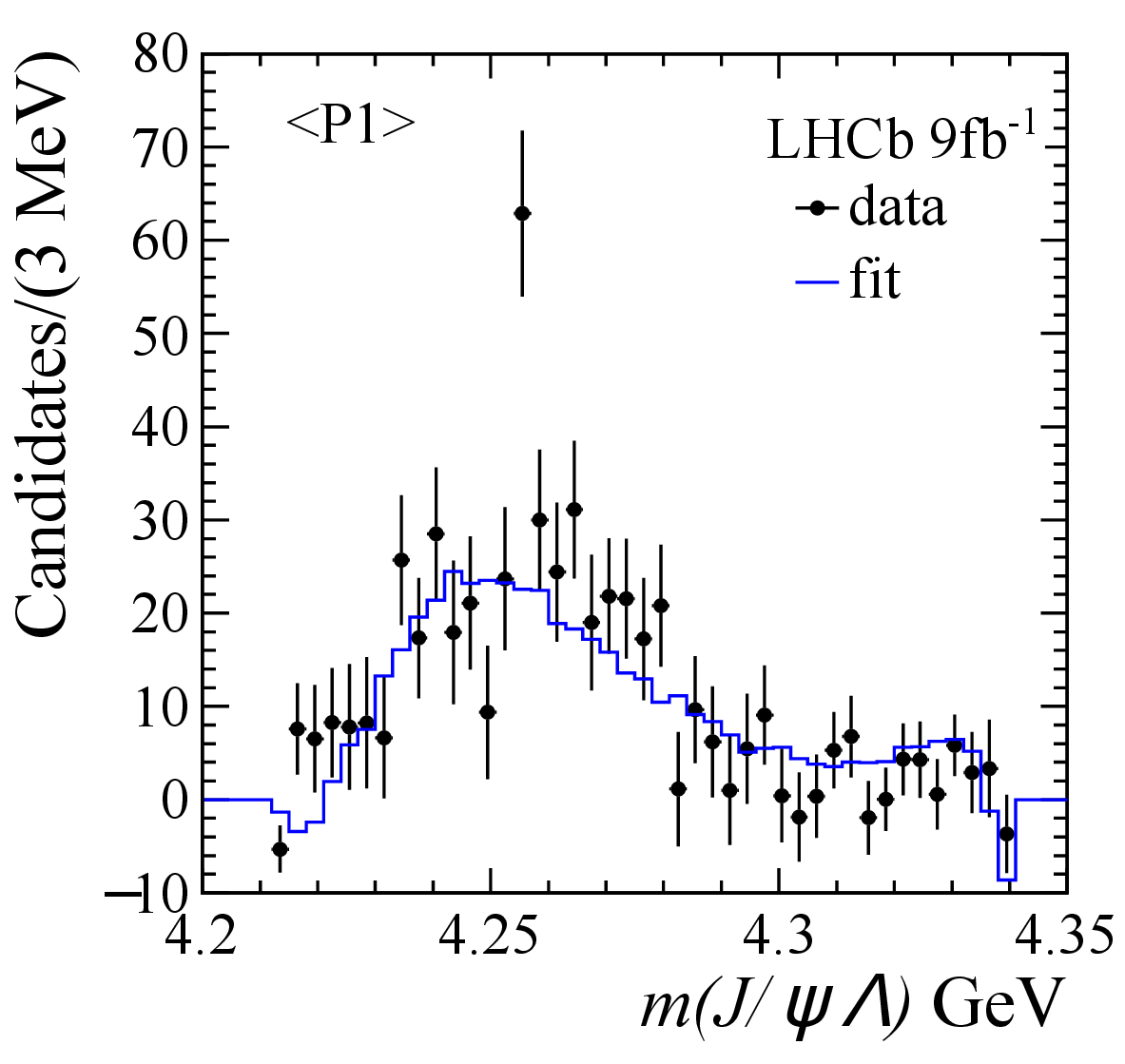}
\includegraphics[width=0.32\textwidth]{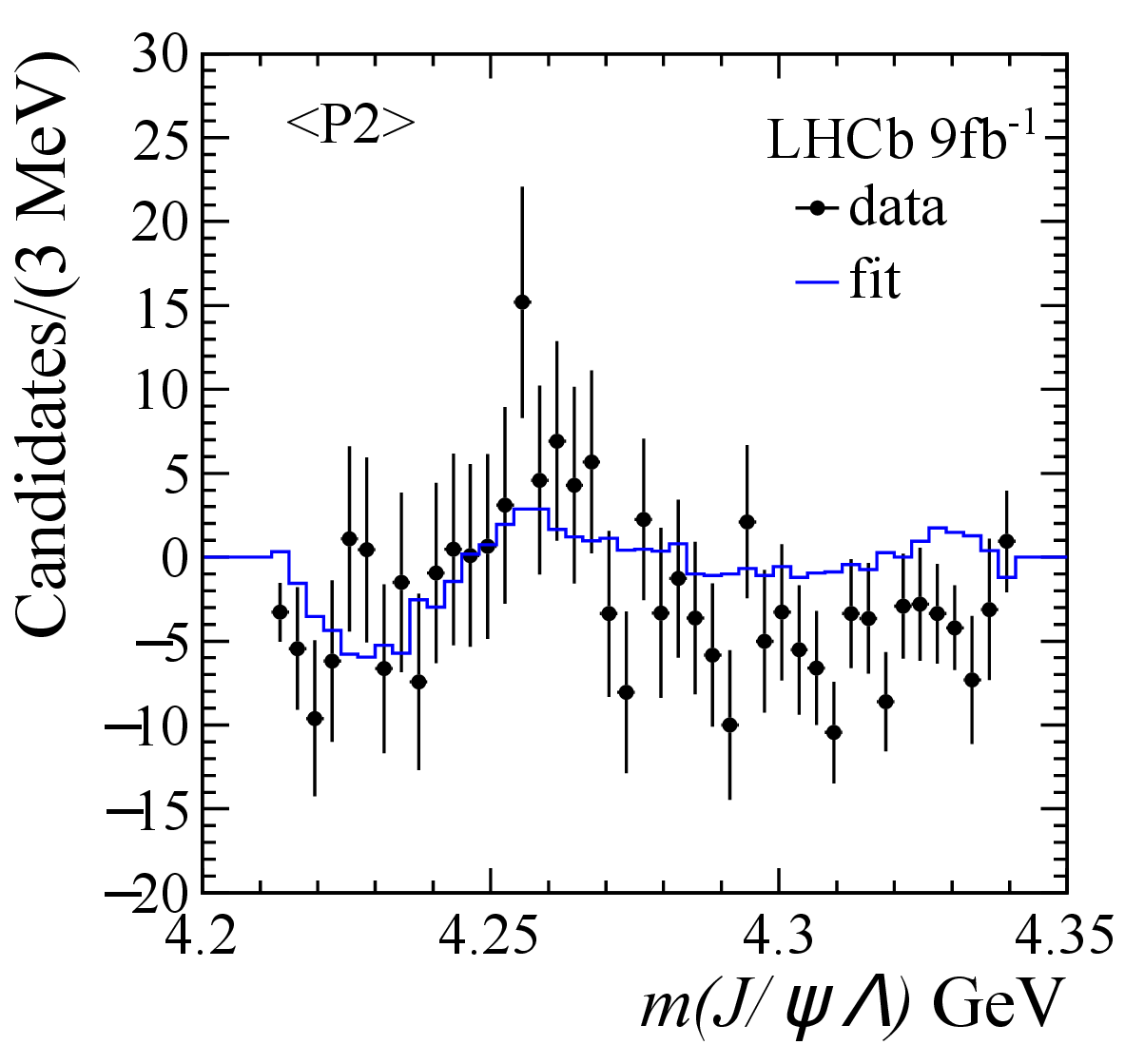}\\
\includegraphics[width=0.32\textwidth]{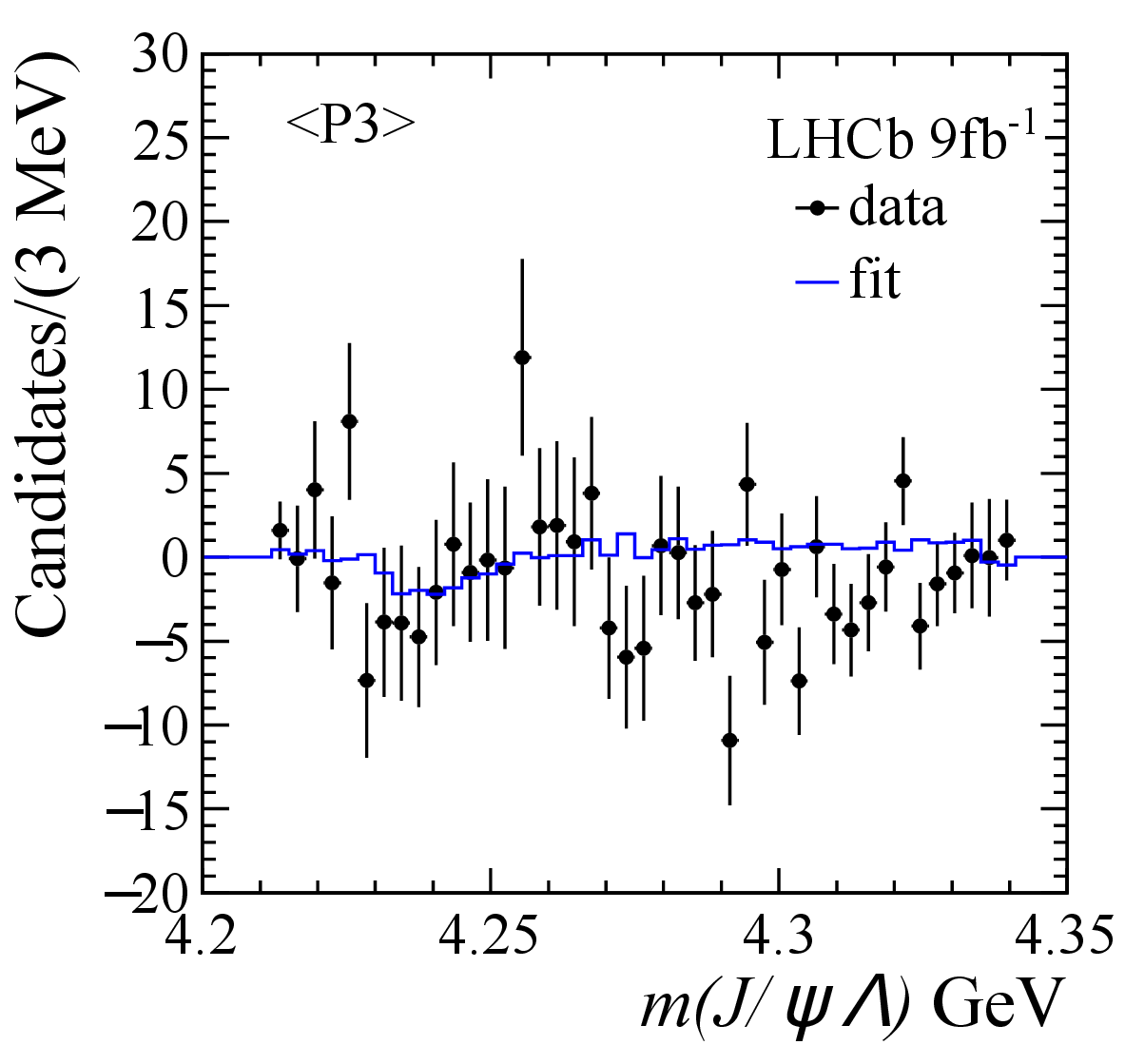}
\includegraphics[width=0.32\textwidth]{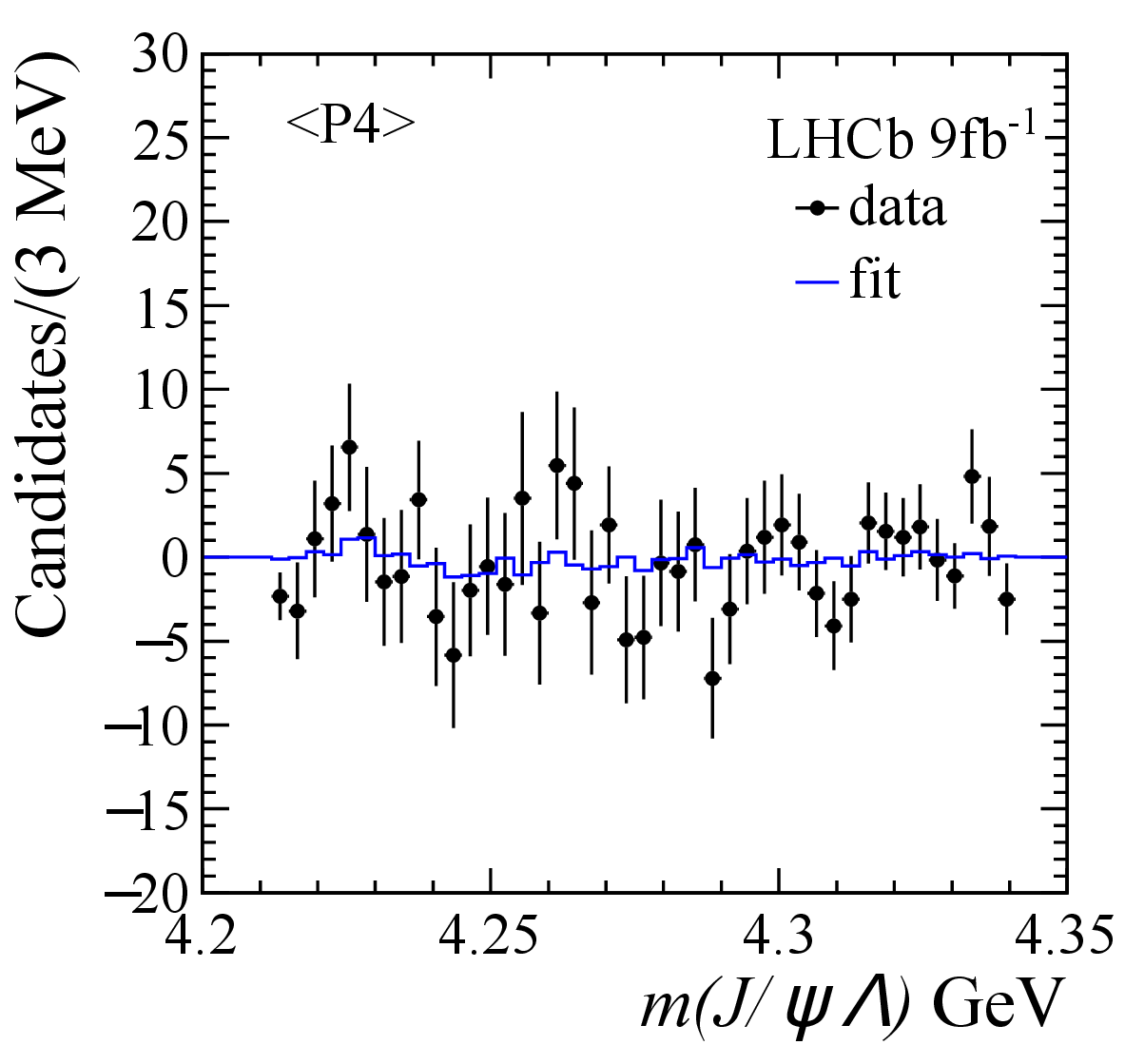}
\includegraphics[width=0.32\textwidth]{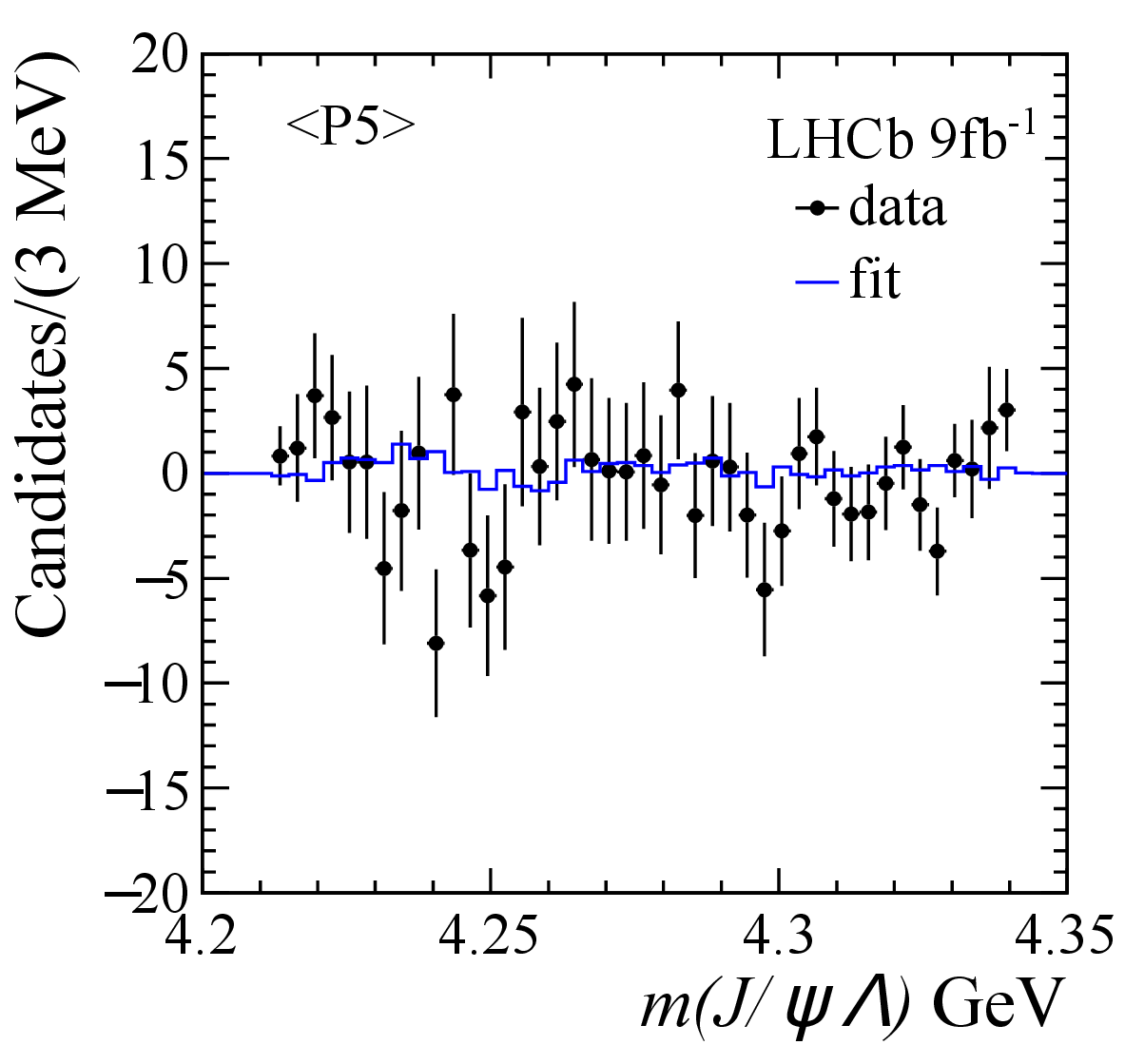}
\caption{\small\label{app:fig:momentsPL}${P}_{\psi s}^{\Lz}(4338)^0$ helicity angular moments as a function of $m(\jpsi\Lz)$ invariant mass. The black points represent the data while the blue line is the nominal model.}
\end{figure}

The angular moments are shown in Fig.~\ref{app:fig:momentsPL}, up to order 5, as a function of the $m(\jpsi\Lz)$ invariant mass distribution. They show a good agreement between the data and the nominal model. 

\section{Efficiency corrected and background subtracted distributions}
\begin{figure}[ht]
\centering
\small
\includegraphics[width=0.4\textwidth]{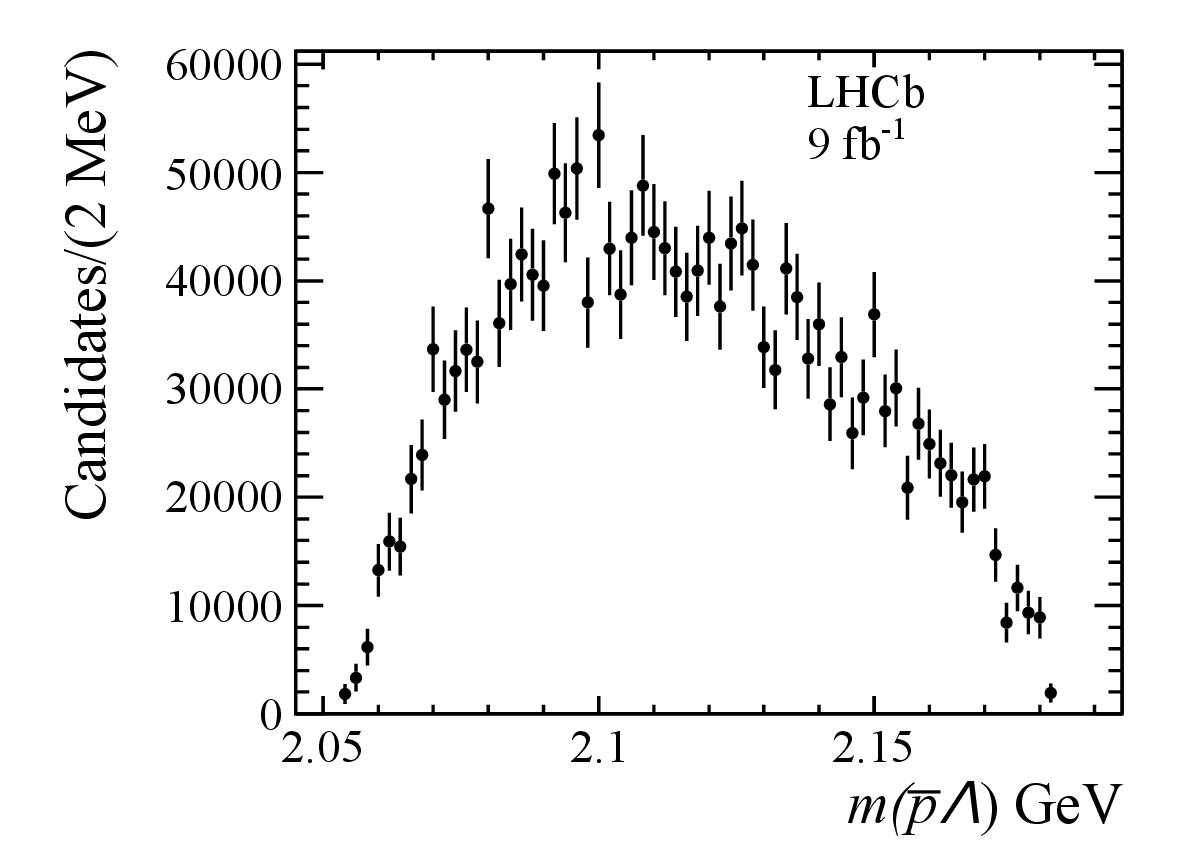}
\includegraphics[width=0.4\textwidth]{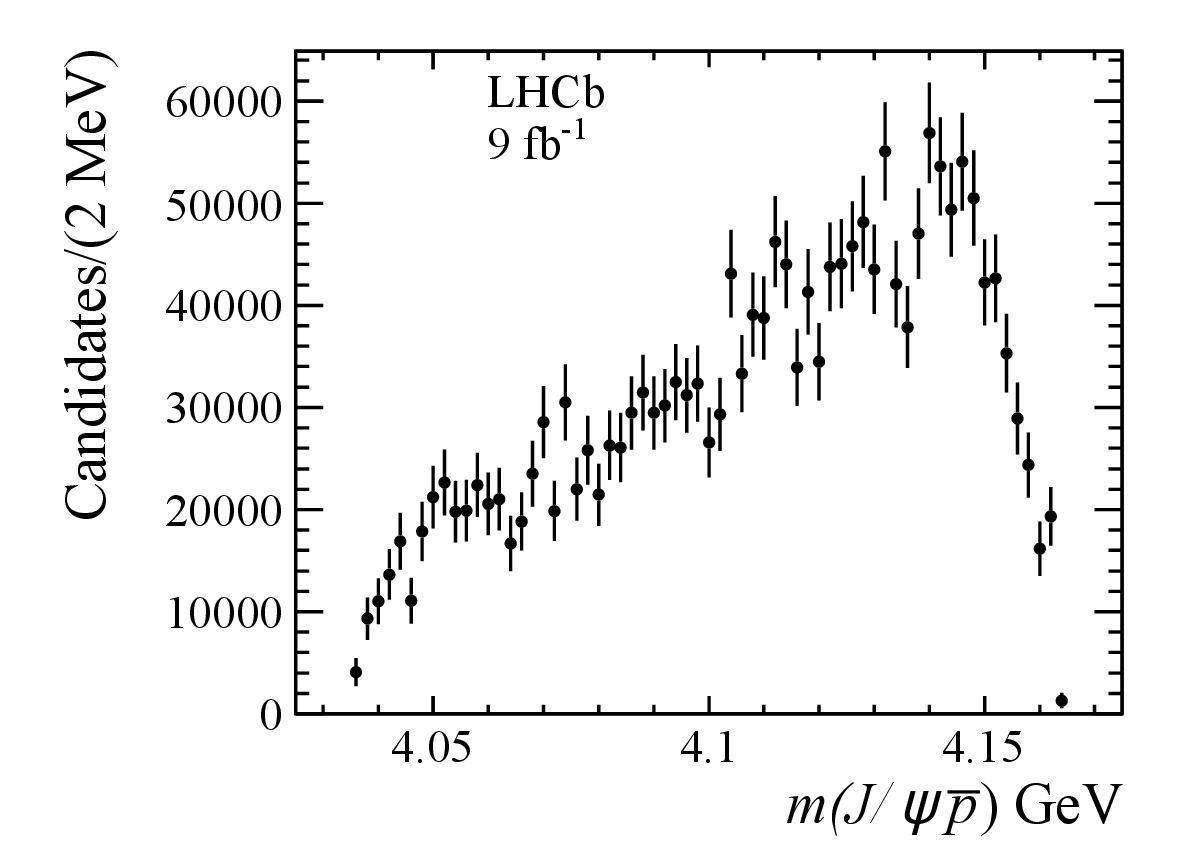}\\
\includegraphics[width=0.4\textwidth]{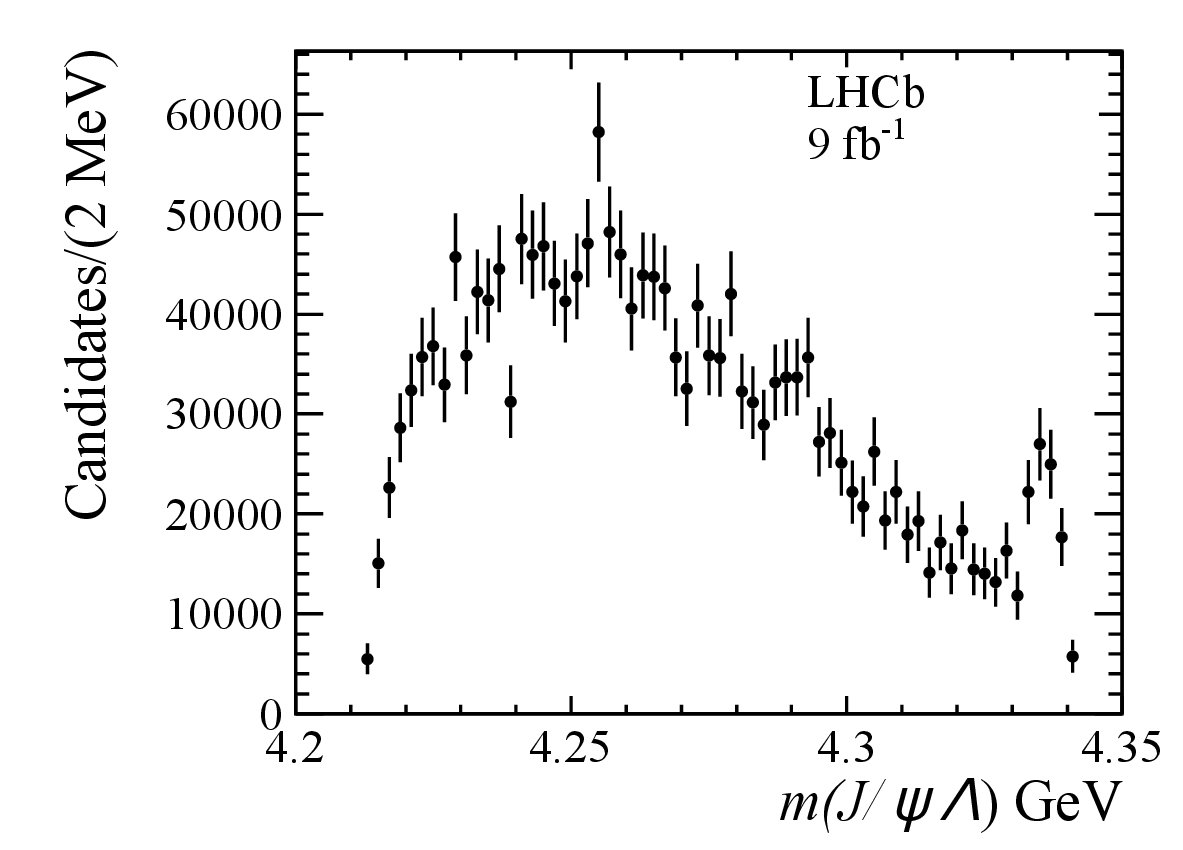}
\includegraphics[width=0.4\textwidth]{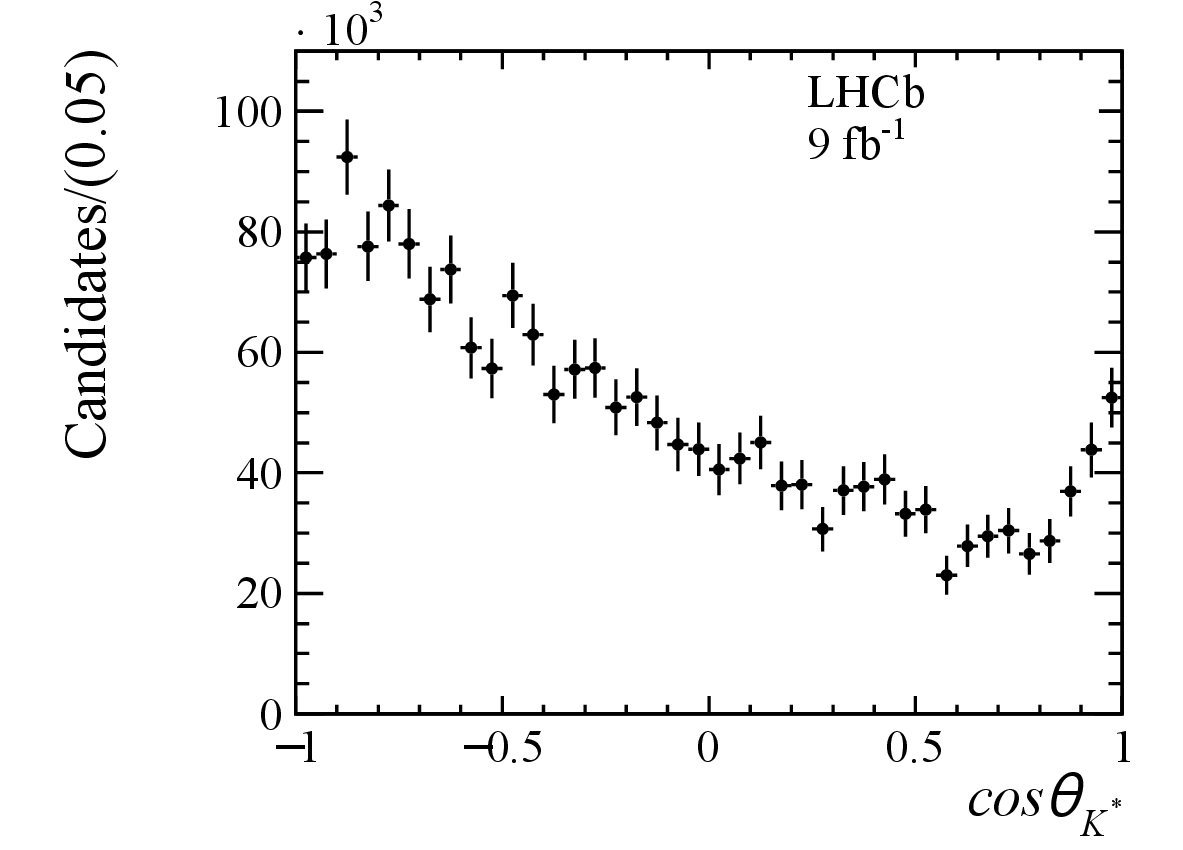}
\caption{Efficiency corrected and background subtracted distributions for $m(\antiproton\Lambda)$, $m(J/\psi \antiproton)$, $m(J/\psi\Lambda)$ and $\cos{\theta_{K^*}}$. }
\label{fig:app:effcorr}
\end{figure}

The data are assigned weights to account for the efficiency and to subtract the background using the \sPlot technique. 
The efficiency corrected data distributions of $m(\antiproton\Lz)$, $m(\jpsi\antiproton)$, $m(\jpsi\Lz)$ and $\cos{\theta_{K^*}}$ are shown in~\figurename~\ref{fig:app:effcorr}.
There is a sign difference between this $\cos{\theta_{K^*}}$ definition and the one from CMS \cite{CMS:2019kbn}.

\section{Invariant mass fit for the \Bm mass measurement}
\begin{figure}[t]
  \centering
  \includegraphics[width=0.5\textwidth]{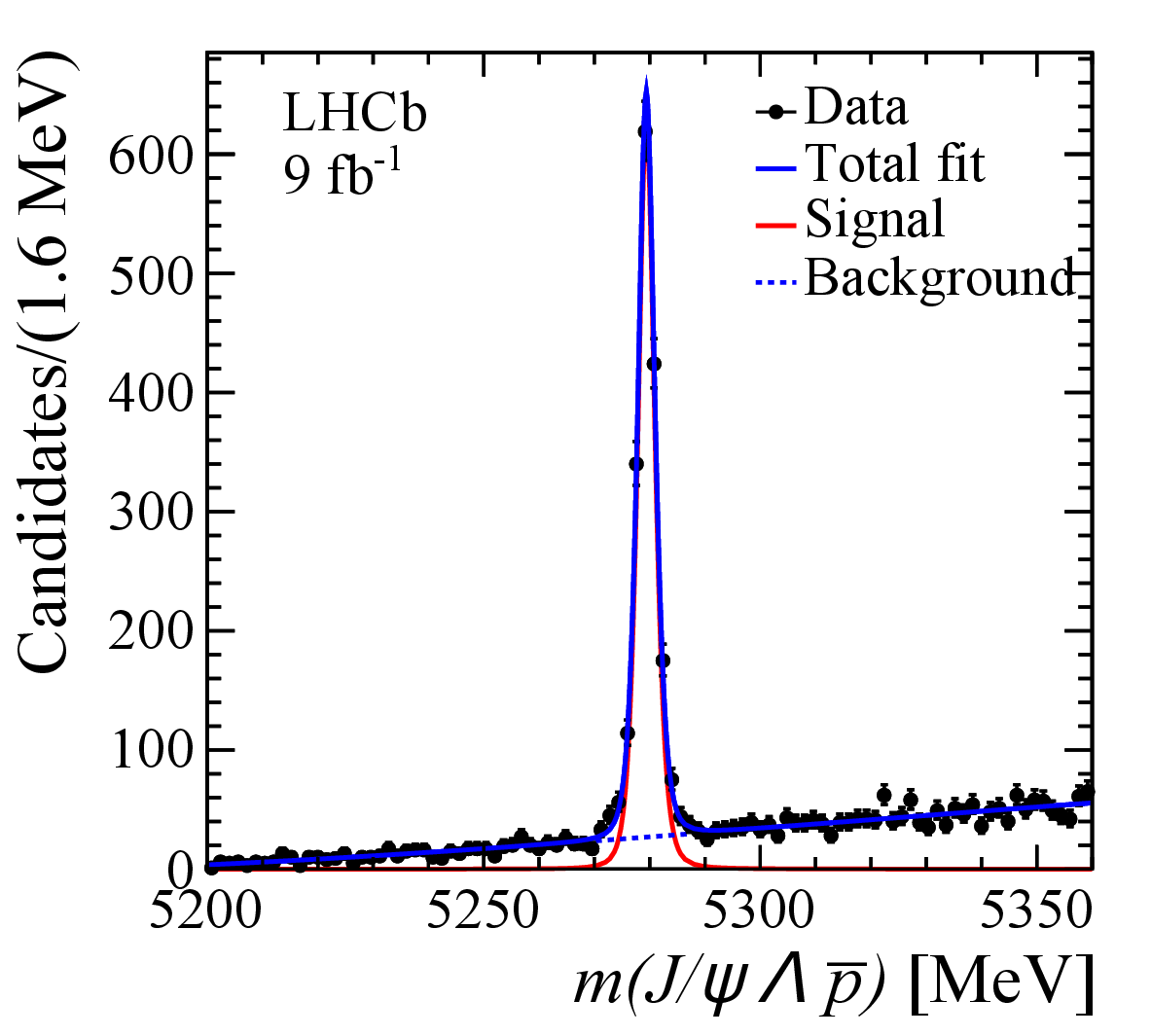}
  \caption{Invariant mass distribution of the \jpsi \Lz \antiproton candidates reconstructued with \Lz baryons in the long category only. This dataset in used for the measurement of the \Bm meson mass. The data are overlaid with the results of the fit.}
  \label{fig:fit_mass_long}
\end{figure}

Fig.~\ref{fig:fit_mass_long} shows the invariant mass fit used to extract the \Bm mass measurement. The signal yield with \Lz baryons in the long category  is $1670\pm40$, which amounts to 36\% of the total candidates. 

%% file: Authorship_LHCb-PAPER-2022-031.tex
\centerline
{\large\bf LHCb collaboration}
\begin
{flushleft}
\small
R.~Aaij$^{32}$\lhcborcid{0000-0003-0533-1952},
A.S.W.~Abdelmotteleb$^{50}$\lhcborcid{0000-0001-7905-0542},
C.~Abellan~Beteta$^{44}$,
F.~Abudin{\'e}n$^{50}$\lhcborcid{0000-0002-6737-3528},
T.~Ackernley$^{54}$\lhcborcid{0000-0002-5951-3498},
B.~Adeva$^{40}$\lhcborcid{0000-0001-9756-3712},
M.~Adinolfi$^{48}$\lhcborcid{0000-0002-1326-1264},
P.~Adlarson$^{77}$\lhcborcid{0000-0001-6280-3851},
H.~Afsharnia$^{9}$,
C.~Agapopoulou$^{13}$\lhcborcid{0000-0002-2368-0147}
% ...

{\footnotesize \it

% ...
$^{9}$Universit{\'e} Clermont Auvergne, CNRS/IN2P3, LPC, Clermont-Ferrand, France\\
% ...
$^{13}$LPNHE, Sorbonne Universit{\'e}, Paris Diderot Sorbonne Paris Cit{\'e}, CNRS/IN2P3, Paris, France\\
% ...
$^{32}$Nikhef National Institute for Subatomic Physics, Amsterdam, Netherlands\\
% ...
$^{40}$Instituto Galego de F{\'\i}sica de Altas Enerx{\'\i}as (IGFAE), Universidade de Santiago de Compostela, Santiago de Compostela, Spain\\
% ...
$^{44}$Physik-Institut, Universit{\"a}t Z{\"u}rich, Z{\"u}rich, Switzerland\\
% ...
$^{48}$H.H. Wills Physics Laboratory, University of Bristol, Bristol, United Kingdom\\
% ...
$^{50}$Department of Physics, University of Warwick, Coventry, United Kingdom\\
% ...
$^{53}$School of Physics and Astronomy, University of Glasgow, Glasgow, United Kingdom\\
$^{54}$Oliver Lodge Laboratory, University of Liverpool, Liverpool, United Kingdom\\
% ...
$^{77}$Department of Physics and Astronomy, Uppsala University, Uppsala, Sweden, associated to $^{53}$\\
% ...
}
\end{flushleft}